\documentclass[12pt,fullpage]{article}
\usepackage{eurosym}
\usepackage{doublespace}
\usepackage{latexsym}
\usepackage{graphicx}
\usepackage{natbib}
\usepackage{lineno}

  \tiny \normalsize
\newcommand{\fourfigv}[4]
{ \hbox to\hsize{\hss
    \includegraphics[scale=0.75,angle=0]{#1}\qquad
    \hskip -.3truein
    \includegraphics[scale=0.75,angle=0]{#2}
    \hss}
\hbox to\hsize{\hss
    \includegraphics[scale=0.75,angle=0]{#3}\qquad
    \hskip -0.3truein
    \includegraphics[scale=0.75,angle=0]{#4}
    \hss}
  \vskip -0.4truein
}

\input{tcilatex}
\begin{document}

\title{Rejection Odds and Rejection Ratios:\ A Proposal for Statistical
Practice in Testing Hypotheses}
\author{M.J. Bayarri \\
\textit{Universitat de Val\`{e}ncia} \and Daniel J. Benjamin \\
\textit{University of Southern California} \and James O. Berger \\
\textit{Duke University} \and Thomas M. Sellke\thanks{{\small Authors are
listed alphabetically. We were greatly saddened by the death of Susie
Bayarri during the preparation of this paper. We are grateful to
Samantha Cunningham and Derek
Lougee for excellent research assistance and to Jonathan Beauchamp,
James Choi, Sander Greenland, Joris Mulder, and Eric-Jan Wagenmakers for
valuable comments. This research was supported by the National Science
Foundation, under grants DMS-1007773, DMS-1407775, and BCS-1521855. E-mails:
daniel.benjamin@gmail.com, berger@stat.duke.edu, tsellke@purdue.edu.}} \\
\textit{Purdue University}}
\date{December 22, 2015}
\maketitle

\abstract{Much of science is (rightly or wrongly) driven by hypothesis testing. Even in
situations where the hypothesis testing paradigm is correct, the common practice of
basing inferences solely on $p$-values has been under intense criticism for over 50 years.
We propose, as an alternative, the use of the odds of a correct rejection of the null hypothesis to incorrect
rejection. Both pre-experimental versions (involving the power and Type I error)
and post-experimental versions (depending on the actual data) are considered.
Implementations are provided that range from depending only on the $p$-value to
consideration of full Bayesian analysis. A surprise is that all implementations --- even
the full Bayesian analysis --- have complete frequentist justification. Versions of our proposal
can be implemented that require only minor modifications to existing practices yet
overcome some of their most severe shortcomings. }

\section{Introduction}

In recent years, many sciences --- including experimental psychology --- have
been embarrassed by a growing number of reports that many findings do not
replicate. While a variety of factors contribute to this state of affairs, a
major part of the problem is that conventional statistical methods, when
applied to standard research designs in psychology and many other sciences,
are too likely to reject the null hypothesis and therefore generate an
unintentionally high rate of false positives. A number of alternative
statistical methods have been proposed, including several in this special
issue, and we are sympathetic to many of these proposals. In particular
we are highly sympathetic to efforts to wean the scientific community
away from an over-reliance on hypothesis testing, with utilization
of often-more-relevant estimation and prediction techniques.

Our goal in this paper is more modest in scope: we propose
a range of modifications --- several \emph{relatively minor} --- to existing statistical practice
in hypothesis testing that we believe would immediately fix
some of the most severe shortcomings of current methodology. The minor modifications
would not require any changes in the
statistical tests that are commonly used, and would rely only on the most
basic statistical concepts and tools, such as significance thresholds,
$p$-values, and statistical power. With $p$-values and power calculations in
hand (obtained from standard software in the usual way), the additional
calculations we recommend can be carried out with a calculator.

In developing and justifying these simple modifications of standard methods,
we also discuss additional tools that are available from Bayesian
statistics. While
these can provide considerable additional benefit in a number of settings,
significant improvements in the testing paradigm can be made even without
them.

We study the standard setting of precise hypothesis testing.\footnote{
By \emph{precise hypothesis testing}, we mean that $H_{0}$ is a lower
dimensional subspace of $H_{1}$, as in (\ref{eq.precise}). In particular,
the major problem with $p$-values that is highlighted in this paper is muted
if the hypotheses are, say, $H_{0}:\theta <0$ versus $H_{1}:\theta >0$. As
an example, suppose $\theta $ denotes the difference in mean treatment
effects for cancer treatments A and B:
\par
\begin{itemize}
\item Scenario 1: Treatment A = standard chemotherapy and Treatment B =
standard chemotherapy + steroids. This is a scenario of precise hypothesis
testing, because steroids could be essentially ineffective against cancer,
so that $\theta $ could quite plausibly be essentially zero.
\par
\item Scenario 2: Treatment A = standard chemotherapy and Treatment B = a
new radiation therapy. In this case there is no reason to think that $\theta
$ could be zero, and it would be more appropriate to test $H_{0}:\theta <0$
versus $H_{1}:\theta >0$.
\end{itemize}
\par
See Berger and Mortera (1999) for discussion of these issues.} We can
observe data $\mbox{\boldmath $x$}$ from the density $f(\mbox{\boldmath $x $}
\mid \theta )$. We consider testing
\begin{equation}
H_{0}:\theta =\theta _{0}\quad \mbox{versus}\quad H_{1}:\theta \neq \theta
_{0}\,.  \label{eq.precise}
\end{equation}
Our proposed approach to hypothesis testing is based on consideration of the
odds of correct rejection of $H_{0}$\ to incorrect rejection. This
`rejection odds'\ approach has a dual frequentist/Bayesian interpretation,
and it addresses four acknowledged problems with common practices of
statistical testing:

\begin{enumerate}
\item Failure to incorporate considerations of power into the interpretation
of the evidence.

\item Failure to incorporate considerations of prior probability into the
design of the experiment.

\item Temptation to misinterpret $p$-values in ways that lead to overstating
the evidence against the null hypothesis and in favor of the alternative
hypothesis.

\item Having optional stopping present in the design or running of the
experiment, but ignoring the stopping rule in the analysis.
\end{enumerate}

There are a host of other problems involving testing, such as the fact that
the size of an effect is often much more important than whether an effect
exists, but here we only focus on the testing problem itself. Our
proposal --- developed throughout the paper and summarized in the
conclusion --- is that researchers should report what we call the `pre-experimental
rejection ratio' when presenting their experimental design, and
researchers should report what we call the `post-experimental
rejection ratio' (or Bayes factor) when presenting their experimental results.

In Section 2, we take a pre-experimental perspective: for a given
anticipated effect size and sample size, we discuss the evidentiary impact
of statistical significance, and we consider the problem of choosing the
significance threshold (the region of results that will lead us to reject
$H_{0}$). The (pre-experimental) `rejection ratio'\ $R_{pre}$, the ratio of
statistical power to significance threshold (i.e., the ratio of the
probability of rejecting under $H_{1}$ and $H_{0}$, respectively), is shown
to capture the strength of evidence in the experiment for $H_{1}$ over $H_{0}
$; its use addresses Problem \#1 above.

How much a researcher should believe in $H_{1}$ over $H_{0}$ depends not
only on the rejection ratio but also on the prior odds, the relative prior
probability of $H_{1}$ to $H_{0}$. The `pre-experimental rejection odds,'\
which is the overall odds in favor of $H_{1}$ implied by rejecting $H_{0}$,
is the product of the rejection ratio and the prior odds. When the prior
odds in favor of $H_{1}$ are low, the rejection ratio need to be greater in
order for the experiment to be equally convincing. This line of reasoning,
which addresses Problem \#2, implies that researchers should adopt more
stringent significance thresholds (and generally use larger sample sizes)\
when demonstrating surprising, counterintuitive effects. The logic
underlying the pre-experimental odds suggests that the standard approach in
many sciences (including experimental psychology) --- accepting $H_{1}$
whenever $H_{0}$ is rejected at a conventional 0.05 significance
threshold --- can lead to especially misleading conclusions when power is low
or the prior odds is low.

In Section 3, we turn to a post-experimental perspective: once the
experimental analysis is completed, how strong is the evidence implied by
the observed data? The analog of the pre-experimental odds is the
`post-experimental odds'\ : the prior odds times the Bayes factor. The Bayes
factor is the ratio of the likelihood of the observed data under $H_{1}$ to
its likelihood under $H_{0}$; for consistency in notation (and because of a
surprising frequentist interpretation that is observed for this ratio), we
will often refer to the Bayes factor as the `post-experimental rejection
ratio,'\ $R_{post}$.

Common misinterpretations of the observed $p$-value (Problem \#3) are that it
somehow reflects the error probability in rejecting $H_{0}$ (see Berger, 2003; Berger, Brown, and Wolpert, 1993)
or the related notion that it reflects the likelihood of the observed data
under $H_{0}$. Both are very wrong. For example, it is sometimes incorrectly
said that $p=0.05$ means that there was only a 5\% chance of observing the
data under $H_{0}$. (The correct statement is that $p=0.05$ means that there
was only a 5\% chance of observing a test statistic as extreme or more
extreme as its observed value under $H_{0}$ --- but this correct statement
isn't very useful because we want to know how strong the evidence is, given
that we actually observed the value 0.05.) Given this misinterpretation,
many researchers dramatically overestimate the strength of the experimental
evidence for $H_{1}$ provided by a $p$-value. The Bayes factor has a
straightforward interpretation as the strength of the evidence in favor of
$H_1$ relative to $H_0$, and thus its use can avoid the misinterpretations that
arise from reliance on the $p$-value.

The Bayes factor approach has been resisted by many scientists because of
two perceived obstacles. First, determination of Bayes factors can be
difficult. Second, many are uneasy about the subjective components of
Bayesian inference, and view the familiar frequentist justification of
inference to be much more comforting. The first issue is addressed in
Section 3.2, where we discuss the `Bayes factor bound'\ $1/[-ep\log p]$
(from Vovk, 1993, and Sellke, Bayarri, and Berger, 2001). This bound is the \emph{largest}
Bayes factor in favor of $H_{1}$ that is possible (under reasonable
assumptions). The Bayes factor bound can thus be interpreted as a best-case
scenario for the strength of the evidence in favor of $H_{1}$ that can arise
from a given $p$-value. Even though it favors $H_{1}$ amongst all
(reasonable) Bayesian procedures, it leads to far more conservative
conclusions than the usual misinterpretation of $p$-values; for example, a $p$-value
of 0.05 only represents \emph{at most} $2.5:1$ evidence in favor of
$H_{1}$. The `post-experimental odds bound'\ can then be calculated as the
Bayes factor bound times the prior odds.

In Section 3.3, we address the frequentist concerns about the Bayes factor.
In fact, we show that in our setting, using the Bayes factor is actually a
fully frequentist procedure --- and, indeed, we argue that it is actually
a much better frequentist procedure than that based on the pre-experimental rejection ratio. Our
result that the Bayes factor has a frequentist justification is novel to
this paper, and it is surprising because the Bayes factor depends on the prior
distribution for the effect size under $H_1$. We point out the resolution to this apparent
puzzle: the prior distribution's role is to prioritize where to maximize
power, while the procedure always maintains frequentist error control for
the rejection ratio that is analogous to Type I frequentist error control.

Our result providing a frequentist justification for the Bayes factor helps to unify the pre-experimental odds with
the post-experimental odds. It also provides a bridge between frequentist and Bayesian
approaches to hypothesis testing.

In Section 4, we address the practical question of how to choose the priors
that enter into the calculation of the Bayes factor and post-experimental odds.
We enumerate a range of options. Researchers may prefer one or another of
these options, depending on whether or not they are comfortable with statistical
modeling, and whether or not they are willing to adopt priors that are subjective.
Since the options cover many situations, we argue that there is
really no practical barrier to reporting the Bayes factor, or at least
the Bayes factor bound. For example, reporting the Bayes factor bound
does not require specifying any prior, and it is simple to calculate from
just the $p$-value associated with standard hypothesis tests. While the Bayes
factor bound is not ideal as a summary of the evidence since it is biased
against the null, we believe its reporting would still lead to more
accurate interpretations of the evidence than reporting the $p$-value alone.

In Section 5, we show that the post-experimental odds approach has the
additional advantage that it overcomes the problems that afflict $p$-values
caused by optional stopping in data collection. A common practice is to
collect some data, analyze it, and then collect more data if the results are
not yet statistically significant (John, Loewenstein, and Prelec, 2012).
There is nothing inherently wrong with such an optional stopping strategy;
indeed, it is a sensible procedure when there are competing demands on a
limited subject pool. However, in the presence of optional stopping, it is
well-known that $p$-values calculated in the usual way can be extremely
biased (e.g., Anscombe, 1954). Under the null hypothesis, there is a greater
than 5\% chance of a statistically significant result when there is more
than one opportunity to get lucky. (Indeed, if scientists were given
unlimited research money and allowed to ignore optional stopping, they would
be guaranteed to be able to reject any correct null hypothesis at any Type I
error level!) In contrast, the post-experimental odds approach we
recommend is not susceptible to biasing via optimal stopping (cf. Berger,
1985, and Berger and Berry, 1988).

The reason that the post-experimental odds do not depend on optional
stopping is that the effect of optional stopping on the likelihood of
observing some realization of the data is a multiplicative constant. When
one considers the odds of one model to another, this same constant is present in the
likelihood of each model, and hence it cancels when taking the ratio of the
likelihoods. Thus, the Bayesian interpretation of post-experimental odds is
unaffected by optional stopping. And since the post-experimental odds have
complete frequentist justification, the frequentist who employs them can
also ignore optional stopping.

In Section 6 we put forward our recommendations for statistical practice in
testing. These recommendations can be boiled down to two: researchers should report
the pre-experimental rejection ratio when presenting their experimental design,
and researchers should report the post-experimental rejection ratio when
presenting their results. How exactly these recommendations should be adopted
will vary according to the standard practice in the science.
For some sciences there are nothing but $p$-values; for others, there are
also minimal power considerations; for some
there are sophisticated power considerations; and for some there are full
blown specifications of prior odds ratios and prior distributions. Our
recommendations accommodate all.

Most of the ideas in this paper have precedents in prior work. What we call
the `pre-experimental rejection odds'\ is nearly identical to Wacholder et
al.'s (2004) `false-positive reporting probability,'\ as further developed
by Ioannidis (2005) and the Wellcome Trust Case Control Consortium (2007).
The `post-experimental rejection odds'\ approach is known in the statistics
literature as Bayesian hypothesis testing. For psychology research, there
have been advocates for Bayesian analysis in general (e.g., Kruschke, 2011)
and for Bayesian hypothesis testing in particular (e.g., Wagenmakers et al.,
in press; Masson, 2011). The Bayes factor bound was introduced by
Vovk (1993) and Sellke,
Bayari, and Berger (2001). We view our contribution primarily as presenting
these ideas to the research community in a unified framework and in terms of
actionable changes in statistical practice. As noted above, however, as far
as we know, our frequentist justification for Bayes factors is a new result
and helps to unify the pre- and post-experimental odds approaches.

\section{The Pre-Experimental Odds Approach:\ Incorporating the Anticipated
Effect Size and Prior Odds in Interpretation of Statistical Significance}

Rejecting the null hypothesis at the 0.05 significance threshold is
typically taken to be sufficient evidence to accept the alternative
hypothesis. Such reasoning, however, is erroneous for at least two reasons.

First, what one should conclude from statistical significance depends not
only on the probability of statistical significance under the null
hypothesis --- the significance threshold of 0.05 --- but also on the probability
of statistical significance under the alternative hypothesis --- the power of
the statistical test. Second, the prior odds in favor of the alternative
hypothesis is relevant for the strength of evidence that should be required.
In particular, if the alternative hypothesis would have seemed very unlikely
prior to running the experiment, then stronger evidence should be needed to
accept it. This section develops a more accurate way to interpret the
evidentiary impact of statistical significance that takes into account these
two points.

\subsection{Pre-Experimental Rejection Odds: the Correct Way to Combine Type
I Error and Power}

Recall that we're interested in testing the null hypothesis, $H_{0}:\theta
=0 $, against the alternative hypothesis, $H_{1}:\theta \neq 0$. Both
standard frequentist and Bayesian approaches can be expressed through choice
of a prior density $\pi (\theta )$ of $\theta $ under $H_{1}$. To a
frequentist, this prior distribution represents a `weight function' for power
computation. Often, this weight function is chosen to be a point mass at a
particular anticipated effect size, i.e., the power is simply evaluated at a
fixed value of $\theta $.

Given the test statistic for the planned analysis (for example, the
$t$-statistic), the rejection region $\mathcal{R}$ is the set of values for the
test statistic such that the null hypothesis is said to be `rejected'\ and
the finding is declared to be statistically significant. The `significance
threshold'\ $\alpha $ (the Type I error probability) is the probability
under $H_{0}$\ that the test statistic falls in the rejection region. In
practice, the rejection region is determined by the choice of the
significance threshold, which is fixed, typically at $\alpha =0.05$. Given
the experimental design and planned analysis, the Type I error probability
$\alpha $ pins down a Type II\ error probability\ $\beta (\theta )$ for each
possible value of $\theta $: the probability that the test statistic does \emph{not} fall
in the rejection region when the parameter equals $\theta $. The average
power is $(1-{\bar{\beta}})\equiv \int (1-\beta (\theta ))\pi (\theta
)d\theta $ (which, again, could simply be the power evaluated at a chosen
fixed effect size).

We want to know: If we run the experiment, what are the odds of \emph{correct}
rejection of the null hypothesis to \emph{incorrect} rejection? We call
this quantity the `pre-experimental rejection odds'\ (sometimes dropping the
word `rejection'\ for brevity).\ Given the definition of the Type I error
probability $\alpha $, the probability of incorrectly rejecting $H_{0}$ is
$\pi _{0}\alpha $, where $\pi _{0}$ is the (prior) probability that $H_{0}$
is true. Given the definition of average power $1-{\bar{\beta}}$, the probability of
correctly rejecting $H_{0}$ is $\pi _{1}(1-{\bar{\beta}})$, where $\pi
_{1}=1-\pi _{0}$ is the (prior) probability that $H_{1}$ is true.
The following definition takes the odds that result from these
quantities and slightly
reorganizes the terms to introduce key components of the odds.

\medskip \noindent \textbf{Definition:} The \emph{pre-experimental odds of
correct to incorrect rejection of the null hypothesis\ }is
\begin{eqnarray}
O_{pre} &=&\frac{\pi _{1}}{\pi _{0}}\times \frac{(1-{\bar{\beta}})}{\alpha }
\\
&\equiv &O_{P}\times R_{pre} \\
&\equiv &\mbox{[prior odds of $H_1$ to $H_0$]}\times
\mbox{[(pre-experimental) rejection ratio
of $H_1$ to $H_0$]}\,.  \nonumber
\end{eqnarray}

An alternative definition of the pre-experimental odds provides a Bayesian
perspective:\ $O_{pre}$ could be defined as the odds of $H_{1}$ to $H_{0}$
conditional on the finding being statistically significant: $O_{pre}\equiv
\frac{\Pr \left( H_{1}\mid \mathcal{R}\right) }{\Pr \left( H_{0}\mid
\mathcal{R}\right) }$. Bayes'\ Rule then implies that $\frac{\Pr \left(
H_{1}\mid \mathcal{R}\right) }{\Pr \left( H_{0}\mid \mathcal{R}\right) }=
\frac{\Pr \left( H_{1}\right) }{\Pr \left( H_{0}\right) }\frac{\Pr \left(
\mathcal{R}\mid H_{1}\right) }{\Pr \left( \mathcal{R}\mid H_{0}\right) }=
\frac{\pi _{1}}{\pi _{0}}\times \frac{(1-{\bar{\beta}})}{\alpha }$, as
above. Of course, this would be the Bayesian answer only if the information
available was just that the null hypothesis was rejected (i.e., $p
\mbox{-value }<\mbox{ }\alpha $); as we will see, if the $p$-value is known,
the Bayesian answer will differ.

The fact that $O_{pre}$ arises as the product of the prior odds and
rejection ratio is scientifically useful, in that it separates prior
opinions about the hypotheses (which can greatly vary) from the \emph{
pre-experimental rejection ratio}, $R_{pre}$, provided by the experiment; this
latter is the odds of rejecting the null hypothesis when $H_{1}$ is true to
rejecting the null hypothesis when $H_{0}$ is true.

Figure 1 illustrates how the ratio of power to the significance threshold
(the rejection ratio)\ represents the evidentiary impact of statistical
significance. Under the null hypothesis $H_{0}:\theta =0$, the red curve
shows the probability density function of the estimated effect $\widehat{
\theta }$, assumed to have a normal distribution. The red shaded region in
the right tail shows the one-sided 5\% significance region. (We focus on the
one-sided region merely to simplify the figure.) The area of the red shaded
region, which equals 0.05, is the probability of observing a statistically
significant result under $H_{0}$. The blue curve shows the probability
density function of the estimated effect $\widehat{\theta }$ under a point
alternative hypothesis $H_{1}:\theta =\theta _{1}>0$. The area of the blue
region plus the area of the red region is the probability of observing a
statistically significant result under $H_{1}$, which equals the level of
statistical power. Their ratio, $\frac{\mbox{red area + blue area}}{
\mbox{red area}}=\frac{1-{\bar{\beta}}}{\alpha }$, is the rejection ratio.

%

\begin{figure}[tb]
\vspace{-1em}
\par
\begin{center}
\includegraphics[scale=0.55]{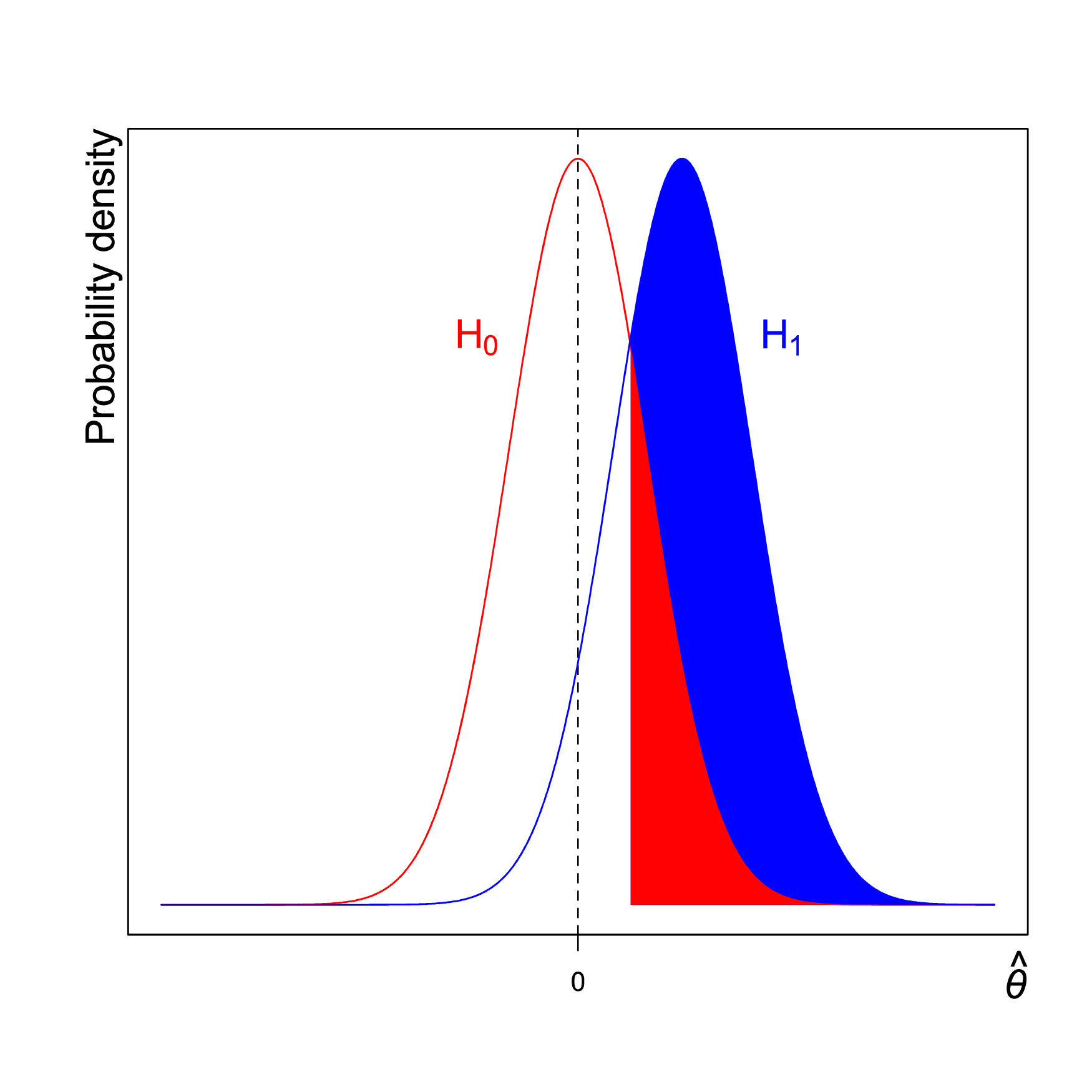} \label{fig:Rplot}
\end{center}
\par
\vspace{-4em}
\caption{The probability density of the observed effect under the null model
(red curve) and a point alternative (blue curve). The area of the red shaded
region, which equals 0.05, is the probability of observing a statistically
significant result under $H_{0}$. The area of the blue region plus the area
of the red region is the probability of observing a statistically
significant result under $H_{1}$, which equals the level of statistical
power.}
\end{figure}

The rejection ratio
takes into account the crucial role of power in understanding the strength
of evidence when rejecting the null hypothesis, and does so in a simple way,
reducing the evidence to a single number. Table 1 shows this crucial role.
For example, in a low powered study with power equal to only 0.25, $\alpha
=0.05$ results in rejection ratio of only $5:1$, which hardly inspires
confidence in the rejection.
\begin{table}[h]
\label{tab.odds1}
\par
\begin{center}
{\small
\begin{tabular}{|r|l|l|l|l|l||l|l|l|l|l|}
\hline
average power ${\bar{\beta}}$ & $0.05$ & $0.25$ & $0.50$ & $0.75$ & $1.0$ & $%
0.01$ & $0.25$ & $0.50$ & $0.75$ & $1.0$ \\ \hline
type I error $\alpha $ & $0.05$ & $0.05$ & $0.05$ & $0.05$ & $0.05$ & $0.01$
& $0.01$ & $0.01$ & $0.01$ & $0.01$ \\ \hline
rejection ratio $R_{pre}$ & $1$ & $5$ & $10$ & $15$ & $20$ & $1$ & $25$ & $%
50 $ & $75$ & $100$ \\ \hline
\end{tabular}
}
\end{center}
\caption{(Pre-experimental) rejection ratios for various Type I errors and
powers.}
\end{table}
Researchers certainly understand that calculating power prior to running an
experiment is valuable in order to evaluate whether the experiment is
sufficiently likely to `work'\ to be worth running in the first place. Once
an effect is found to be statistically significant, however, there is a
common but faulty intuition that statistical power is no longer relevant.%
\footnote{
This line of reasoning --- `if the evidence is unlikely under
the the null hypothesis, then the evidence favors the alternative
hypothesis' --- may be a version of
`confirmatory bias' (see, e.g., Fischhoff and Beyth-Marom,
1983, pp.247-248). The person making the judgment (in this case, the
researcher) is considering only the consistency or inconsistency of the
evidence with respect to one of the hypotheses, rather than with respect to
both of two possible hypotheses.}

Calculating the rejection ratio requires knowing the power of the
statistical test, which in turn requires specifying an anticipated effect
size or more generally a prior distribution over effect sizes $\pi (\theta )$.
Of course, choosing an anticipated effect size can be tricky and sometimes
controversial; see Gelman and Carlin (2014) for some helpful discussion
of how to use external information to guide the choice. We
advocate erring on the conservative size (i.e., assuming `too small'\ an
effect size) because many of the relevant biases in human judgment push in
the direction of assuming too large an effect size. For example, researchers
may be subject to a `focusing illusion'\ (Schkade and Kahneman, 1998),
exaggerating the role of the hypothesized mechanism due to not thinking
about other mechanisms that also matter. As another example, since obtaining
a smaller sample size is usually less costly than a larger sample size,
researchers may wishfully convince themselves that the effect size is large
enough to justify the smaller sample.

We also caution researchers against uncritically relying on meta-analyses
for determining the anticipated effect size. There are three reasons. First,
there may be publication bias in the literature due to the well known `file
drawer problem'\ (Rosenthal, 1979): the experiments that did not find an
effect may not be published and thus may be omitted from the meta-analyses,
leading to an upward bias in the estimated effect size. Second and
relatedly, if experiments that find a significant effect are more likely to
be included in the analysis, then the estimated effect size will be biased
upward due to the `winner's curse'\ (e.g., Garner, 2007): conditional on
statistical significance, the effect estimate is biased away from zero
(regardless of the true effect size). Third and finally, if any of the
studies included in the meta-analysis adopted questionable research
practices that inflate the estimated effects (see John, Loewenstein, and
Prelec, 2012) or practices that push estimated effects toward the null (such
as using unusually noisy dependent variables), then the meta-analysis
estimate will be correspondingly biased.\bigskip

\noindent \textbf{Example 1:} \emph{The Effect of Priming Asian Identity on
Delay of Gratification}: In an experiment conducted with Asian-American
undergraduate students, Benjamin, Choi, and Strickland (2010) tested whether
making salient participants' Asian identity increased their willingness to
delay gratification. Participants in the treatment group ($n=37$) were asked
to fill out a questionnaire that asked about their family background.
Participants in the control group ($n=34$) were asked instead to fill out a
questionnaire unrelated to family background. In both groups, after filling
out the questionnaire, participants made a series of choices between a
smaller amount of money to be received sooner (either today or in 1 week)
and a larger amount of money to be received later (either in 1 week or in 2
weeks). The research question was how often participants in the treatment
group made the patient choice relative to participants in the control group.

What was the pre-experimental rejection ratio $R_{pre}$ for this
experiment?\footnote{
To be clear, while we conduct this calculation here, and post-experimental
calculations in section 3.2 below, the authors did not report such
calculations.} A conservative anticipated effect size may be $d=0.26$, where
`Cohen's $d$'\ is the difference in means across treatment groups in
standard-deviation units (a common effect-size measure for meta-analyses in
psychology). This value was the average effect size reported in a
meta-analysis of explicit semantic priming effects (Lucas, 2000), such as
the effect of seeing the word `doctor' \ on
the speed of subsequent judgments about the conceptually related word
`nurse.' \ Given that hypothetical effect
size and the actual sample sizes, the power of the experiment was 0.19. Thus
$R_{pre}$ was only $\frac{0.19}{0.05}=3.8$.\bigskip

What rejection ratio should be considered acceptable? One answer is implicit
in the conventions for significance threshold (0.05) and acceptable power
(0.80). In that case, the rejection ratio is $16:1$. While choosing a
threshold for an `acceptable'\ rejection ratio is somewhat arbitrary, to
maintain continuity with existing conventions, we will adopt a threshold of $
16:1$ for ordinary circumstances (but we will discuss circumstances when a
different threshold is warranted in the next subsection).

Planned sample sizes should be sufficient to ensure an adequate rejection
ratio. If the rejection ratio of the planned experiment is too small, then
the experiment is not worth running because even a statistically significant
finding does not provide much information. The directive to plan for a
rejection ratio of $16:1$ will often be equivalent to the usual directive to
plan for 80\% power.

Unfortunately, current norms in many sciences often lead to \emph{much} less
than 80\% power. Indeed, low power appears to be a
problem in a range of disciplines, including psychology (Vankov, Bowers, and
Munaf\`{o}, 2014; Cohen, 1988), neuroscience (Button et al., 2013),\ and experimental
economics (Zhang and Ortmann, 2013).
To illustrate, we
use Richard, Bond, and Stokes-Zoota's (2003) review of meta-analyses across
a wide range of research topics in social psychology. Averaged across
research areas, they estimate a `typical'\ effect size of $r=0.21$, where $r$
is the Pearson product--moment correlation coefficient between the dependent
variable and a treatment indicator (another common effect-size measure for
meta-analyses in psychology). Given this effect size, Table 2 shows
statistical power and the rejection ratio at the 0.05 significance threshold
for an experiment conducted with different sample sizes. For simplicity, we
assume that each observation is drawn from a normal distribution with unit
variance. For the control group, the mean is 0, while for the treatment
group, the mean is 0.21. We assume that the treatment and control group each
have a sample size of $n$.
\begin{table}[h]
\begin{center}
{\small
\begin{tabular}{|r|l|l|l|l|l|l|l|l|l|l|}
\hline
per-condition $n$ & $10$ & $20$ & $30$ & $40$ & $50$ & $100$ & $150$ & $200$
& $250$ & $280$ \\ \hline
power ${\bar{\beta}}$ & $0.12$ & $0.16$ & $0.20$ & $0.24$ & $0.28$ & $0.44$
& $0.57$ & $0.68$ & $0.76$ & $0.80$ \\ \hline
type I error $\alpha $ & $0.05$ & $0.05$ & $0.05$ & $0.05$ & $0.05$ & $0.05$
& $0.05$ & $0.05$ & $0.05$ & $0.05$ \\ \hline
rejection ratio $R_{pre}$ & $2.4$ & $3.3$ & $4.1$ & $4.8$ & $5.5$ & $8.7$ & $%
11.4$ & $13.5$ & $15.2$ & $16.0$ \\ \hline
\end{tabular}%
}
\end{center}
\caption{For a fixed effect size of $r=0.21$, this table shows the statistical power and
rejection ratio at the 0.05 significance threshold for an experiment
conducted with different sample sizes. }
\end{table}

In some areas of psychology, typical sample sizes are as small as
$n=20$ participants per condition, and in many fields, typical sample sizes
are smaller than 50 per condition. Given an effect size of $r=0.21$,
however, 280 participants per condition are needed for 80\% power. Of
course, there are substantial differences in typical effect sizes across
research areas, and in any particular case, the power calculations should be
suited to the appropriate anticipated effect size.

\subsection{Setting the Significance Threshold $\protect \alpha $}

By convention, in psychological research and many other sciences, the
statistical significance threshold $\alpha $ is almost always set equal to
0.05. Thinking about the pre-experimental odds sheds light on why 0.05 is
often \emph{not} an appropriate significance threshold, and it provides a
framework for determining a more appropriate level for $\alpha $. (While we
highly recommend scientists consider tailoring the significance threshold
to reflect the prior odds,
this subsection can be skipped, without loss of continuity,
by those who do not want to consider prior odds.)

Recall that the pre-experimental odds depend not only on the rejection
ratio, but also the prior odds: $O_{pre}=\frac{\pi _{1}}{\pi _{0}}\times
\frac{(1-{\bar{\beta}})}{\alpha }$. A statistical test that has rejection
ratio of $16:1 $ has pre-experimental odds of $16:1$ only if, prior to the
experiment, $H_{1}$ and $H_{0}$\ were considered equally likely to be true.
If $H_{1}$ has a much lower prior probability than $H_{0}$, say the prior
odds are less than $1:16$, then the pre-experimental odds are less than one
even if $p<0.05 $.

In fact, since power can never exceed 100\%, when the significance threshold
is 0.05, the largest possible rejection ratio is $1:0.05$ = $20:1$.
Therefore, when $\alpha =0.05$, if the prior odds are less than $1:20$, the
null hypothesis remains more likely than the alternative hypothesis even
when the result is in the rejection region.\bigskip

\noindent \textbf{Example 2:} \emph{Evidence for Parapsychological Phenomena}:
In a controversial paper, Bem (2011) presented evidence in favor of
parapsychological phenomena from 9 experiments with over 1,000 participants.
There have been many criticisms of this paper. Wagenmakers, Wetzels,
Borsboom, and van der Maas's (2011) `Problem 2'\ can be understood in terms
of the pre-experimental odds framework presented here. While it is of course
highly speculative to put a prior probability on the existence of
parapsychological phenomena, for illustrative purposes Wagenmakers et al.
assume $\pi _{1}=10^{-20}$ (and $\pi _{0}=1-\pi _{1}$). With such a
skeptical prior probability, what is the evidentiary impact of statistical
significance at the 0.05 threshold? Not much. Since the rejection ratio is
bounded above by $20:1$, the pre-experimental odds can be at most
$\frac{10^{-20}}{1-10^{-20}}\frac{20}{1}\approx 2\times 10^{-19}$.

\smallskip
When the prior odds are low, the significance threshold needs to be made
more stringent in order for statistical significance to constitute
convincing enough evidence against the null hypothesis.\bigskip

\noindent \textbf{Example 3:} \emph{Genome-Wide Association Studies}: Early
genomic epidemiological studies had a low replication rate because they were
conducting hypothesis tests at standard significance thresholds. In 2007, a
very influential paper by the Wellcome Trust Case Control Consortium
proposed instead a cutoff of $p<5\times 10^{-7}$. The argument for this was
a pre-experimental odds argument. Using the earlier notation, they argued
that $O_{P}=\frac{1}{100,000}$, assumed that $(1-{\bar{\beta}})=0.5$, and
wanted pre-experimental odds of $10:1$ in order to claim a discovery.
Solving for $\alpha $ yields $\alpha =5\times 10^{-7}$. Using this
criterion, the paper reported 21 genome/disease associations, virtually all
of which have been replicated.

Subsequent work tightened the significance threshold further, and the
current convention for `genome-wide significance'\ is $\alpha =5\times
10^{-8}$. Genome-wide association studies using this threshold have
continued to accumulate a growing number of robust findings (Visscher,
Brown, McCarthy, and Yang, 2012; Benjamin et al., 2014).\bigskip

Of course, adopting a more stringent significance threshold than 0.05 will
mean that, for a given anticipated effect size, attaining an adequate level
of statistical power will require larger sample sizes --- perhaps \emph{much}
larger sample sizes. Indeed, recent genome-wide association studies that
focus on complex traits (influenced by many genetic variants of small
effect), such as height (Wood et al., 2014), obesity (Locke et al., 2015),
schizophrenia (Ripke et al., 2014), and educational attainment (Rietveld et
al., 2013), have used sample sizes of over 100,000 individuals.

We suspect that our examples of parapsychological phenomena and genome-wide
association studies are extreme within the realm of experimental psychology;
most domains will not have prior odds quite so stacked in favor of the null
hypothesis. Nonetheless, we also suspect that many domains of experimental
psychology should adopt significance thresholds more stringent than 0.05 and
should generally feature studies with larger sample sizes than are currently
standard.

\section{The Post-Experimental Rejection Odds Approach: Finding the
Rejection Odds Corresponding to the Observed Data}

While pre-experimental rejection odds (and the pre-experimental rejection
ratio) are relevant prior to seeing the data, their use after seeing the
data has been rightly criticized by many (e.g., Lucke, 2009). After all, the
pre-experimental rejection ratio for an $\alpha =0.05$-level study might be
$16:1$, but should a researcher report $16:1$ regardless of whether $p=0.05$
or $p=0.000001$?

One of the main attractions in reporting $p$-values is that they
measure the strength of evidence against the null hypothesis in a
way that is data-dependent. But reliance on the $p$-value tempts researchers into
erroneous interpretations. For example, $p=0.01$ does \emph{not} mean that
the observed data had a 1\% chance of occurring under the null hypothesis;
the correct statement is that, under the null hypothesis, there is a
1\% chance of a test statistic as extreme or more extreme than what
was observed. And when correctly interpreted, the $p$-value has some
unappealing properties. For example, it measures the likelihood that
the data would be more extreme than they were, rather than being a measure
of the data that were actually observed. The $p=0.01$ also focuses
exclusively on the null hypothesis, rather than directly addressing the (usually more
interesting) question of how strongly the evidence supports the alternative
hypothesis relative to the null hypothesis.

In this section, we present the post-experimental rejection odds. Like their
pre-experimental cousin discussed in the last section, the post-experimental
odds focus on the strength of evidence for the alternative hypothesis relative
to the null hypothesis. However, the post-experimental odds are data-dependent
and have a straightforward interpretation as the relative probability of the
hypotheses given the observed data.

\subsection{Post-Experimental Odds}

The \emph{post-experimental rejection odds} (also
called posterior odds) of $H_{1}$ to $H_{0}$ is the probability density
of $H_{1}$ given the data divided by the probability density
of $H_{0}$ given the data. These odds are derived via Bayes Rule
analogously to the Bayesian derivation of the pre-experimental rejection odds,
except conditioning the observed data \mbox{\boldmath $x$} rather than
on the rejection region $\mathcal{R}$. The post-experimental odds are given by
\begin{eqnarray}
O_{post}(\mbox{\boldmath $x$}) &=&\frac{\pi _{1}}{\pi _{0}}\times \frac{m(
\mbox{\boldmath $x$})}{f(\mbox{\boldmath $x $}\mid \theta _{0})}  \nonumber
\\
&\equiv &O_{P}\times R_{post}(\mbox{\boldmath $x$}),
\end{eqnarray}
where $R_{post}(\mbox{\boldmath $x$})$ is the \emph{post-experimental
rejection ratio} (more commonly called the Bayes factor or weighted
likelihood ratio) of $H_{1}$ to $H_{0}$, and
\begin{equation}
m(\mbox{\boldmath $x$})=\int_{\{ \theta \neq \theta _{0}\}}f(
\mbox{\boldmath $x
$}\mid \theta )\pi (\theta )d\theta  \label{eq.marginal}
\end{equation}
is the marginal likelihood of the data under the prior $\pi (\theta )$ for $
\theta $ under the alternative hypothesis $H_{1}$. Figure 2 illustrates,
in the same context as in Figure 1, observed data in the rejection region.
The post-experimental rejection ratio is the ratio of the probability
density under $H_1$ to the probability density under $H_0$. Clearly $R_{post}(
\mbox{\boldmath $x$})$ depends on the actual data $\mbox{\boldmath $x$}$
that is observed.

\begin{figure}[tb]
\vspace{-1em}
\par
\begin{center}
\includegraphics[scale=0.55]{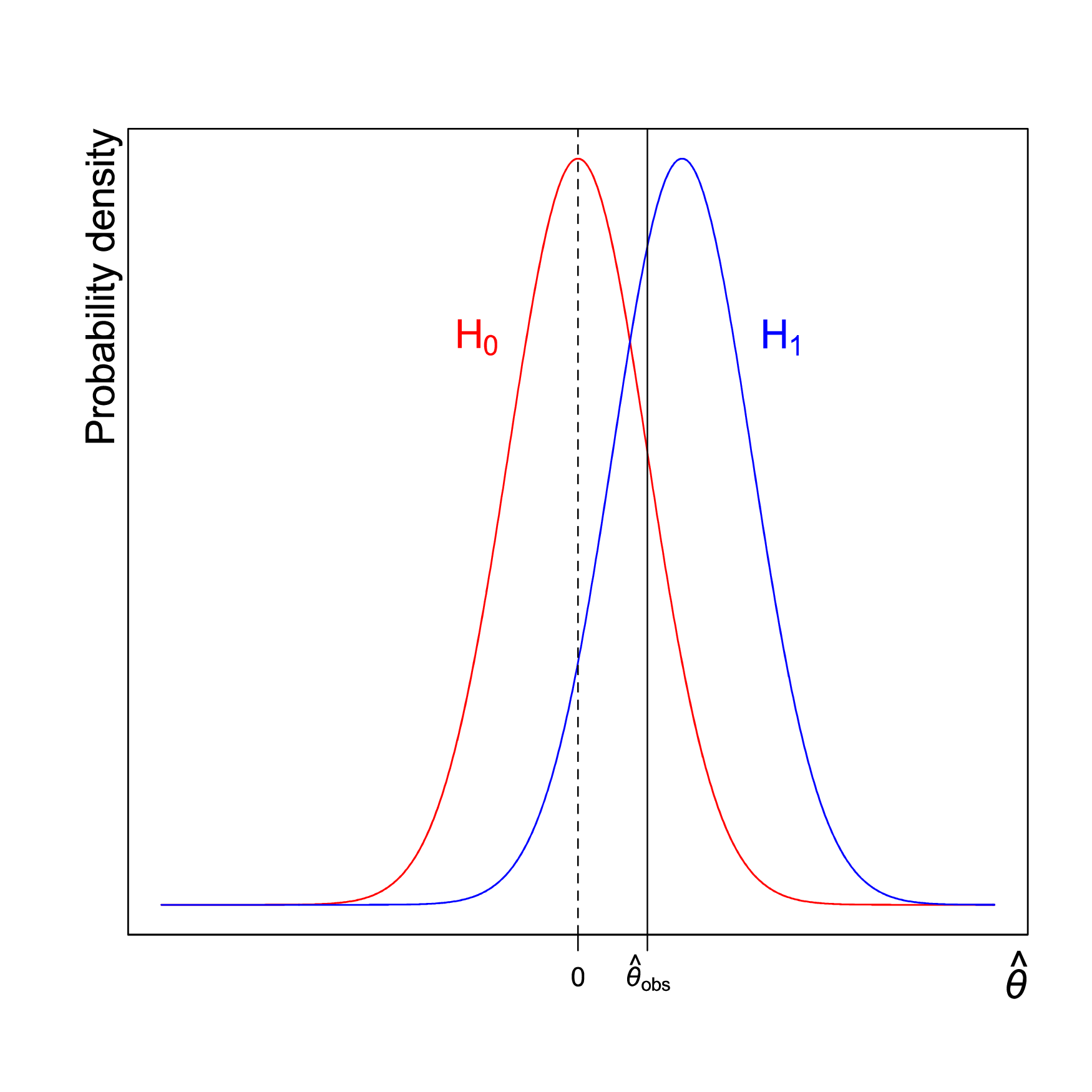} \label{fig:2}
\end{center}
\par
\vspace{-4em}
\caption{The post-experimental rejection ratio (Bayes factor) is the
ratio of the probability density of ${\hat \theta}_{obs}$ under $H_1$ to the
probability density of ${\hat \theta}_{obs}$ under $H_0$.}
\end{figure}

We utilize this standard Bayesian framework to discuss post-experimental
odds, but note that we will be presenting fully frequentist and default
versions of these odds --- i.e., versions that do not require specification of
any prior distributions.\bigskip

\noindent \textbf{Example 4:} \emph{Effectiveness of An AIDS Vaccine}:
Gilbert et al. (2011) reports on a study conducted in Thailand investigating
the effectiveness of a proposed vaccine for HIV. The treatment consisted of
using two previous vaccines, called Alvac and Aidsvax, in sequence, the
second as a `booster' given several weeks after the first. One interesting
feature of the treatment was that neither Alvac nor Aidsvax had exhibited
any efficacy individually in preventing HIV, so many scientists felt that
the the prior odds for success were rather low. But clearly
some scientists felt that the prior odds for success were reasonable (else
the study would not have been done); in any case, to
avoid this debate we focus here on just $
R_{post}(\mbox{\boldmath $x$})$, rather than the post-experimental odds.

A total of 16,395 individuals from the general (not high-risk) population
were involved, with 74 HIV cases being reported from the 8,198 individuals
receiving placebos, and 51 HIV cases reported in the 8,197 individuals
receiving the treatment. The data can be reduced to a $z$-statistic of 2.06,
which will be approximately normally distributed with mean $\theta $, with
the null hypothesis of no treatment effect mapping into $H_{0}:\theta =0$.
If we consider the alternative hypothesis to be $H_{1}:\theta >0$, $z=2.06$
yields a one-sided $p$-value of 0.02.

To compute the pre-experimental rejection ratio, the test was to be done at
the $\alpha =0.05$ level, so $\mathcal{R}=(1.645,\infty )$ would have been
the rejection region for $z$. The researchers calculcated the power of the
test to be $1-\bar{\beta}=0.45$, so $R_{pre}=(1-\bar{\beta})/\alpha =9$.
Thus, pre-experimentally, the rejection ratio was $9:1$, i.e., a rejection
would be nine times more likely to arise under $H_{1}$ than under $H_{0}$
(assuming prior odds of $1:1$). It is worth emphasizing again that many
misinterpret the $p$-value of 0.02 here as implying $50:1$ odds in favor of $
H_{1}$, certainly not supported by the pre-experimental rejection ratio, and
even less supported by the actual data, as we will see.

Writing it as a function of the $z$-statistic, the Bayes factor of $H_{1}$
to $H_{0}$ in this example is
\[
R_{post}(z)=\frac{\int_{0}^{\infty }\frac{1}{\sqrt{2\pi }}e^{-(z-\theta
)^{2}/2}\pi (\theta )d\theta }{\frac{1}{\sqrt{2\pi }}e^{-(z-0)^{2}/2}}\,,
\]
and depends on the choice of the prior distribution $\pi (\theta )$ under $
H_{1}$. Here are three interesting choices of $\pi (\theta )$ and the
resulting post-experimental rejection odds:

\begin{itemize}
\item Analysis of power considerations in designing the study suggested a
`study team'\ prior\footnote{This prior distribution was determined for {\em vaccine efficacy} (VE), which
is the percentage of individuals for which the vaccine prevents infection, rather than the simpler
parameter $\theta$ used in the illustration herein. The study team prior density on VE was
uniform from -20\% (the vaccine could be harmful) to +60\%.}, utilization of which results in ${R_{post}(2.06)}=4.0$.

\item The nonincreasing prior \emph{most favorable} to $H_{1}$ is $\pi
(\theta )=\mbox{Uniform}(0,2.95)$, and yields ${R_{post}(2.06)}=5.63$. (It
is natural to restrict prior distributions to be nonincreasing away from the
null hypothesis, in that there was no scientific reason, based on previous
studies, to expect any biological effect whatsoever.)

\item For \emph{any} prior, $R_{post}(2.06) \leq 8.35$, the latter achieved
by a prior that places a point mass at the maximum likelihood estimator of $
\theta$ (Edwards, Lindman, and Savage, 1963).
\end{itemize}

Thus the pre-experimental rejection ratio of $9:1$ does not accurately
represent what the data says. Odds of $4:1$ or $5:1$ in favor of $H_{1}$ are
indicated when $z=2.06$, and $9:1$ is not possible for {\em any} choice of the
prior distribution of $\theta $.

\subsection{A Simple Bound on the Post-experimental Rejection Ratio (Bayes
Factor), Requiring Only the $p$-value}

\label{sec:plogp}

In this section, for continuity with the statistical literature we draw on,
we revert to using the Bayes factor language. Calculating Bayes factors
requires some statistical modeling (as illustrated in the above example),
which may be a substantial departure from the norm in some research
communities. Indeed, one reason for the ubiquitous reporting of
$p$-values is the simplicity therein; for example, one need not worry about
power, prior odds, or prior distributions. We have argued strongly that
consideration of these additional features is of great importance in
hypothesis testing but we do not want the lack of such
consideration to justify the continued current practice with $p$-values. It
would therefore be useful to have a way of obtaining something like the
Bayes factor using only the $p$-value. In addition, having such a method
would enable assessing the strength of evidence from historical published
studies, from which it is often not possible to reconstruct power or prior
information.

Here is the key result relating the Bayes factor to the $p$-value:

\medskip \noindent \textbf{\emph{Result 1:}} Under quite general conditions,
if the $p$-value is \emph{proper} (i.e., $p(\mbox{\boldmath $x$})$ has a
uniform distribution under the null hypothesis) and if $p\leq 1/e\approx
0.37 $, then
\begin{equation}
R_{post}(\mbox{\boldmath $x$})\leq \frac{1}{-ep\log p}\,.  \label{eq.bound}
\end{equation}
Note that this bound depends on the data only through the $p$-value
(and note that the logarithm is the natural log).
The bound was first developed in Vovk (1993) under the assumption that the
distribution of $p$-values under the alternative is in the class of $
Beta(1,b)$ distributions. The result was generalized by \citet{SellBayaBerg:12}, who showed
that it holds under a natural assumption on the hazard rate of the
distribution under the alternative. Roughly, the assumption (which is
implicitly a condition on $\pi (\theta )$) is that, under the alternative distribution,
$Pr(p < \frac{1}{2}p_0 \mid p <p_0)$
increases as $p_0 \rightarrow 0$, so that the distribution of $p$ under the alternative
concentrates more and more around 0 as one moves close to zero.
The bound was further studied in Sellke (2012),
who showed it to be accurate under the assumptions made in a wide variety of
common hypothesis-testing scenarios involving two-sided testing. For one-sided
precise hypothesis testing (e.g. $H_0: \theta=0$ versus $H_1: \theta >0$),
Sellke (2012) showed that the bound no longer need strictly hold, but that
any deviations from the bound tend to be minor.

Although the result provides merely an upper bound on the Bayes factor, it is
nonetheless highly useful: we know that the post-experimental rejection ratio can never
be larger than this bound. Table 3 shows the value of the Bayes factor bound
for $p $-values ranging from the conventional `suggestive significance'\ threshold
of 0.1 to the `genome-wide significance'\ thresholds mentioned in Example 3.

\begin{table}[h]
\begin{center}
\begin{tabular}{|r|l|l|l|l|l|l|l|l|l|}
\hline
$p$ & $0.1$ & $0.05$ & $0.01$ & $0.005$ & $0.001$ & $0.0001$ & $0.00001$ & $
5\times 10^{-7}$ & $5\times 10^{-8}$ \\ \hline
$\frac{1}{-ep\log (p)}$ & $1.60$ & $2.44$ & $8.13$ & $13.9$ & $52.9$ & $400$
& $3226$ & $2.0\times 10^{5}$ & $2.3\times 10^{6}$ \\ \hline
\end{tabular}
\end{center}
\caption{Values of the Bayes factor upper bound for various values of the $p$
-value.}
\end{table}

An important implication of these calculations is that results that just reach conventional levels
of significance do not actually provide very strong evidence against the null
hypothesis. A $p $-value of $0.05$ could correspond to a post-experimental rejection ratio of at
most $2.44:1$. A $p $-value of $0.01$ -- often considered `highly significant' -- could
correspond to a post-experimental rejection ratio of at
most $8.13:1$, which falls well short of our standard of $16:1$.

Although we do not push it in this paper, one could argue that the
significance threshold should be chosen so that any result achieving statistical
significance constitutes strong evidence against the null hypothesis --- especially
since, in practice, researchers are tempted to interpret significant results
in this way. In that case, the research community may want to change the
standard significance threshold from 0.05 to 0.005, a $p$-value
that yields a bound on the odds that is close to $16:1$ rejection odds.
Interestingly, this was also the significance threshold proposed in Johnson
(2013); the reasoning therein was quite different, but it is telling that
various attempts to interpret the meaning of $p$-values are converging
to similar conclusions. (And as discussed above, this threshold should be made
more stringent if the probability of $H_{1}$ is small.)

If the bound is not large, then rejecting the null hypothesis does \emph{not}
strongly suggest that the alternative hypothesis is true --- but because it is
an upper bound, its interpretation when it \emph{is} large is less clear.
The following example illustrates.\bigskip

\medskip \noindent \textbf{Example 1 (continued):} \emph{The Effect of
Priming Asian Identity on Delay of Gratification}: Recall from above that,
for Benjamin, Choi, and Strickland's (2010) test of whether making salient
participants' Asian identity increased their willingness to delay
gratification, the rejection ratio were only $3.8:1$. But given what they
found, how strong is the evidence against the null hypothesis?

Benjamin et al. reported that participants in the treatment group made the
patient choice 87\% of the time, compared with 74\% of the time in the
control group ($t(69)=3.43$, $p=0.001$). The Bayes factor bound is thus $
\frac{1}{-e \times 0.001 \times \log \left( 0.001\right) }=52.9$. We can
conclude that $R_{post}(\mbox{\boldmath $x$})\leq 52.9$, but this, by
itself, does not allow for a strong claim of significance because it is an
upper bound. Indeed, we argue below that, in low-powered studies such as
this one, the Bayes factor bound is likely to be far too high.\bigskip

In such situations, one could, of course, compute the Bayes factor $R_{post}(
\mbox{\boldmath $x$})$ explicitly (and again, this will be shown to have
complete frequentist justification). But if this cannot be done, we argue
for use of the bound as the post-experimental rejection ratio, for two
reasons.

The first reason is simply that $1/[-ep\log p]$ is much smaller than $1/p$,
so reporting the former is much better than just reporting $p$ and then
misinterpreting $1/p$ as being the odds. The second reason is that there is
some empirical evidence that indicates that Bayes factors frequently are
reasonably close to the bounds. In particular, Figure 3 displays $p$-values
versus the reciprocal of estimated Bayes factors, $1/[R_{post}]$, across
studies in a range of scientific fields (these data are from
Ioannidis, 2008, and Elgersma and Green, 2009). These Bayes factors
have a corresponding {\em lower bound} equal to $[-e p \log p]$, shown as a
hatched curve in all four panels. It can be seen that many of the
estimated results lie fairly close to this lower bound.

\begin{figure}[tbp]
{ \hbox to\hsize{\hss
    \includegraphics[scale=0.75,angle=0]{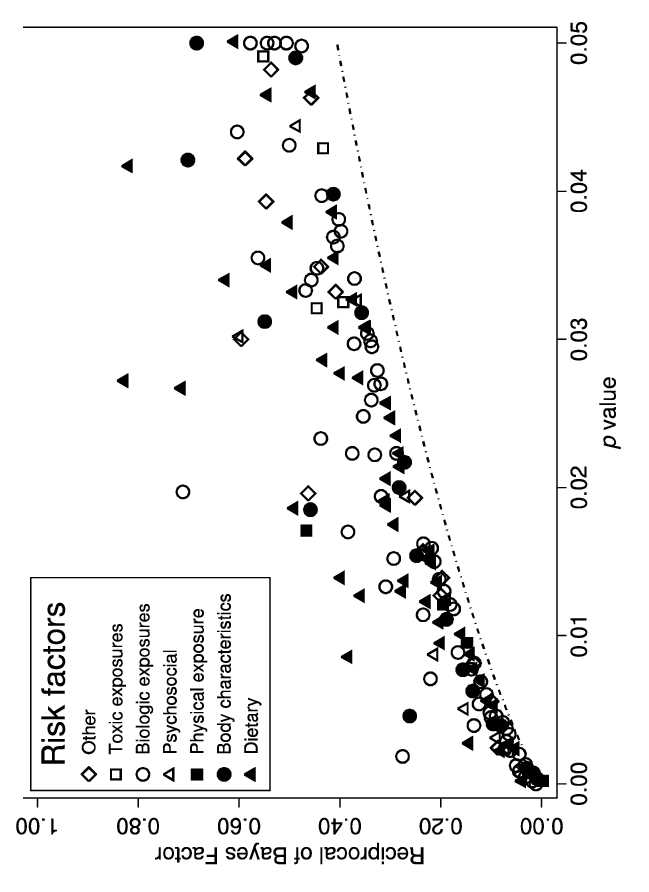}\qquad
    \hskip -.3truein
    \includegraphics[scale=0.75,angle=0]{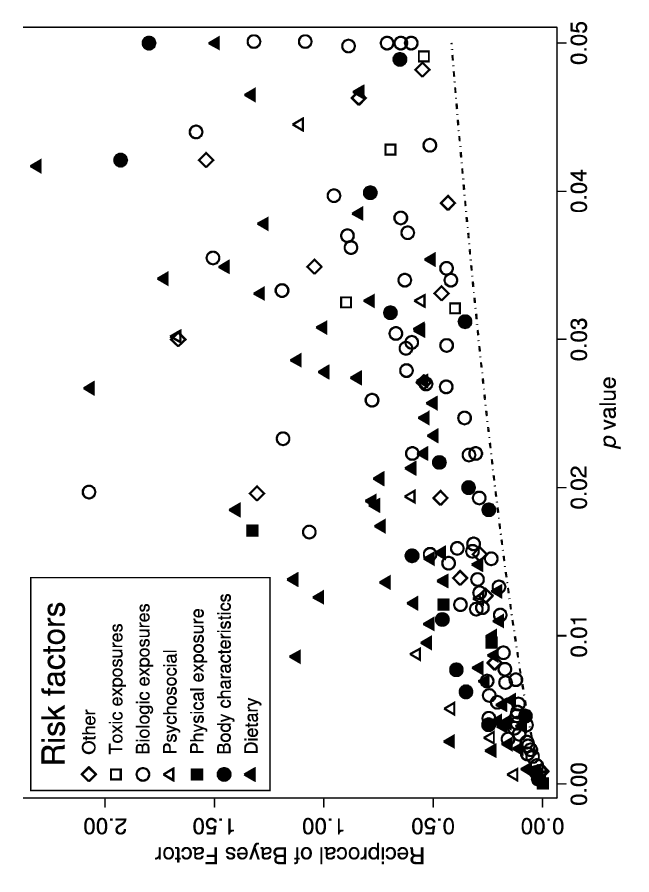}
    \hss}
\hbox to\hsize{\hss
    \includegraphics[scale=0.75,angle=0]{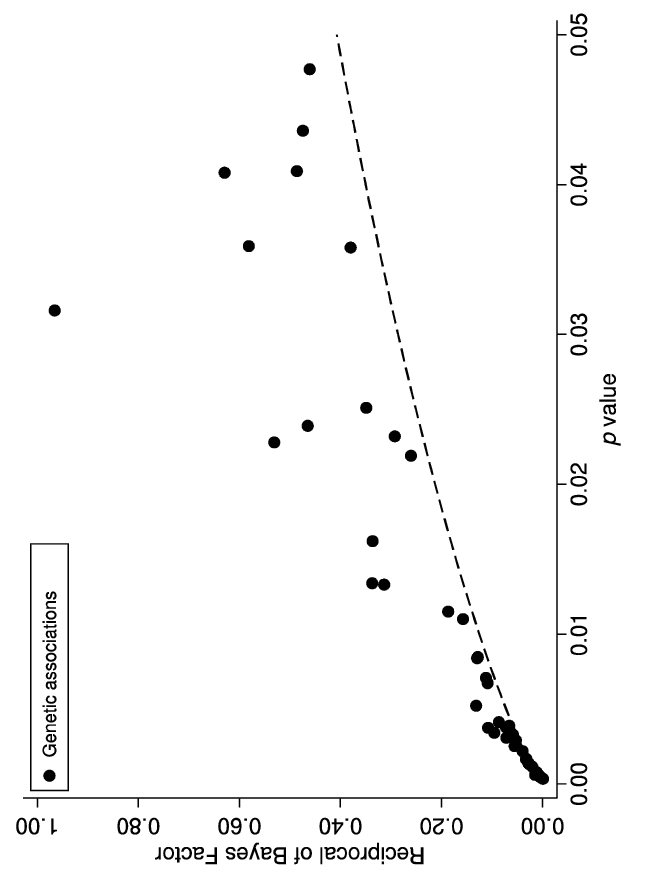}\qquad
    \hskip -0.3truein
    \includegraphics[scale=0.75,angle=0]{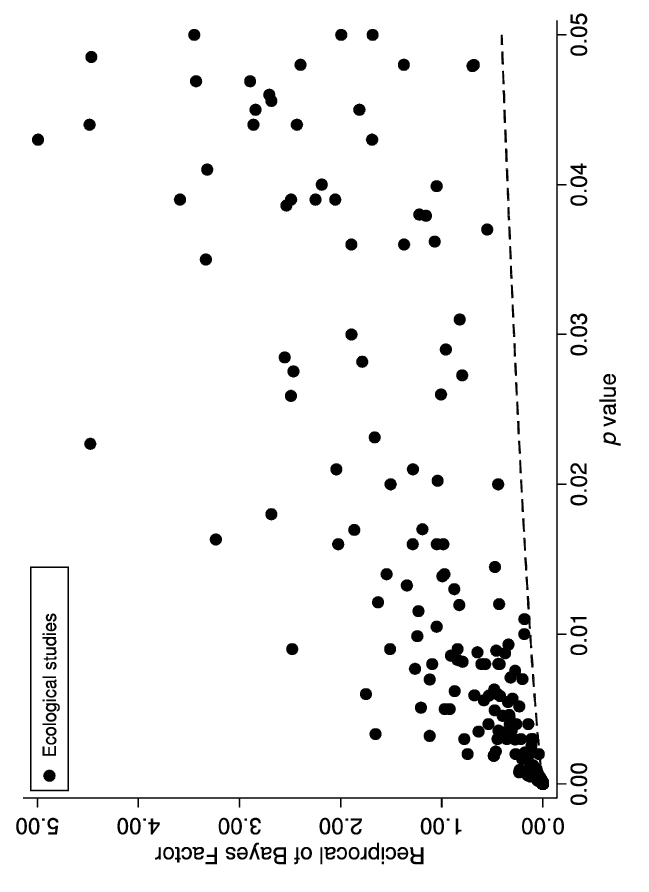}
    \hss
  }
}
\caption{These figures show the relationship between the $p$-value and
the reciprocal of the estimated Bayes factor, $1/R_{post}$, for results
that were statistically significant $(p < 0.05)$ across a range of
scientific fields. In each figure, the dashed line shows the
reciprocal of the Bayes factor bound $(-e p \log p)$, which is a lower
bound for the reciprocal of the Bayes factor. The top two panels
graph $1/R_{post}$ versus the $p$-value for 272
epidemiological studies; $1/R_{post}$ is estimated assuming a
relative risk under the alternative of 1.65 in the first panel
 and 4.48 in the second panel. The lower left panel
 gives the same for 50 genetic associations; $1/R_{post}$ is estimated assuming a
relative risk under the alternative of 1.44 (the median observed
across the genetic associations). The last panel graphs
$1/R_{post}$ versus $p$-values for 202 articles published in
the journal {\em Ecology} in 2009; $1/R_{post}$ is estimated assuming a
standardized effect size under the alternative drawn from
$Uniform[-6,6]$. This distribution was used in previous work (Sellke et
al., 2001), which found that Bayes factor calculations were fairly
robust to alternative plausible distributions. The first three panels
are Figures 1-3 from Ioannidis (2008). The last panel is an edited version of
Figure 4a from Elgersma and Green (2011); of the 308 articles included
in the data for that paper, we dropped 6 articles with a $p$-value
greater than 0.05, and we dropped 6 additional articles with $1/R_{post}$
 greater than 5 in order to
make the figure more readable.}
\end{figure}

While these empirical findings are promising, the situation for low powered
studies can be considerably worse, as shown in the following
example.

\newpage
\noindent \textbf{Example 5:}
 \emph{The Bayes Factor Bound And Power}: An
extreme example\footnote{Our example involves a hypothesis test about the
variance of a normal distribution, even though a hypothesis test about
the mean would be much more standard in applications. The problem with
a hypothesis test about the mean of a normal distribution, such as
$H_0: \mu=0$ versus $H_1: \mu = 0.25$, is that it calls for a one-sided
hypothesis test. As noted above, however, the Bayes factor bound need
not strictly hold for one-sided tests (Sellke, 2012). While a two-sided
test would be common in practical applications, it is unfair to compare
such a non-optimal frequentist procedure with the Bayes factor. The
basic point of our example --- that the Bayes factor bound provides a
better approximation to the Bayes factor in higher powered
studies --- would extend to testing hypotheses about the mean of
a normal distribution.} of the difference that can arise from low power alternatives
is that of observing one observation $X\sim N( 0,\sigma ^{2})$ and testing
\[
H_{0}:\sigma ^{2}=1\quad \mbox{versus}\quad H_{1}:\sigma ^{2}=1.1\,.
\]%
Since the null hypothesis is the usual one, the $p$-value is also just the
usual one, e.g., $p=0.05$ if $x=1.96$.

Here, for a rejection region of the form $\left \vert X\right \vert \geq
1.96 $, the power is just $0.0617$, so $R_{pre}$ is only 1.233. The Bayes
factor (here, just the likelihood ratio between the hypotheses), for a given $x$, is
\[
R_{post}(x)=(0.953)e^{x^{2}/22}\,.
\]%
Table 4 shows the huge discrepancy between the strength
of evidence suggested by $p$ and the strength of evidence implied by the Bayes factor, but
also the large discrepancy between the Bayes factor and $1/[-e\,p\, \log p]$%
.
\begin{table}[h]
\begin{center}
\begin{tabular}{|r|l|l|l|l|l|l|l|}
\hline
$x$ & 1.65 & 1.96 & 2.58 & 2.81 & 3.29 & 3.89 & 4.42 \\ \hline
$p$ & $0.1$ & $0.05$ & $0.01$ & $0.005$ & $0.001$ & $0.0001$ & $0.00001$ \\
\hline
$R_{post}(x) $ & $1.079$ & $1.135$ & $1.290$ & $1.365$ & $1.559$ & $1.897$ &
$2.317$ \\ \hline \hline
$1/[-e\,p\, \log p]$ & $1.598$ & $2.456$ & $7.988$ & $13.89$ & $53.25$ & $%
399.4$ & $3195$ \\ \hline
\end{tabular}%
\end{center}
\caption{For Example 5 (a low powered test), there is a large discrepancy between the strength
of evidence suggested by $p$ and the strength of evidence implied by the Bayes factor,
but also a large discrepancy between the Bayes factor and its
upper bound.}
\end{table}

The situation improves considerably for higher powered studies. In testing $%
H_{0}:\sigma ^{2}=1$ versus more `separated'\ alternatives --- in particular
the alternative values $4,9$, and $16$ --- for the same critical region $%
\left \vert X\right \vert \geq 1.96$ and observed $p$-value $p=0.05$, the
pre-experimental rejection ratio is $R_{pre}=6.54,10.27,$ and $12.48$,
respectively, and the Bayes factors and upper bounds are much closer, as is
shown in Table 5.
\begin{table}[h]
\begin{center}
\begin{tabular}[t]{|r|l|l|l|l|l|l|l|}
\hline
$x$ & 1.65 & 1.96 & 2.58 & 2.81 & 3.29 & 3.89 & 4.42 \\ \hline
$p$ & $0.1$ & $0.05$ & $0.01$ & $0.005$ & $0.001$ & $0.0001$ & $0.00001$ \\
\hline
$R_{post}(x), \sigma^2=4$ & $1.388$ & $2.112$ & $6.067$ & $9.659$ & $28.96$
& $145.7$ & $759.8$ \\ \hline
$R_{post}(x), \sigma^2=9$ & $1.118$ & $1.838$ & $6.422$ & $11.14$ & $40.94$
& $277.8$ & $1967$ \\ \hline
$R_{post}(x), \sigma^2=16$ & $.8957$ & $1.513$ & $5.662$ & $10.12$ & $39.94$
& $300.9$ & $2372$ \\ \hline \hline
$1/[-e\,p\, \log p]$ & $1.598$ & $2.456$ & $7.988$ & $13.89$ & $53.25$ & $%
399.4$ & $3195$ \\ \hline
\end{tabular}%
\end{center}
\caption{In Example 5, the discrepancies between the Bayes factor and the
upper bound are considerably reduced for more separated alternatives.}
\end{table}

\subsection{The Surprising Frequentist/Bayesian Synthesis}

The post-experimental odds presented in the previous section was derived as
a Bayesian evaluation of the evidence. Surprisingly, this Bayesian answer is
also a frequentist answer. To clarify this claim, begin by recalling the
frequentist principle.

\medskip \noindent \textbf{Frequentist Principle:} \emph{In repeated
practical use of a statistical procedure, the long-run average actual
accuracy should not be less than (and ideally should equal) the long-run
average reported accuracy.}\footnote{Note that our statement of the
principle refers here to `repeated practical use.' This is in
contrast to stylized textbook statements, which tend to focus on
the fictional case of drawing new samples and re-running the {\em same}
experiment over and over again. As Neyman himself repeatedly pointed
out (see, e.g., Neyman, 1977), the real motivation for the frequentist
theory is to provide a {\em procedure} --- for example, rejecting the
null hypothesis when $p < 0.05$ --- that, if used repeatedly
for {\em different} experiments, would on average have the correct
level of accuracy.}

Here is the key result showing that the post-experimental rejection ratio
(Bayes factor) is a valid frequentist report:

\medskip \noindent \textbf{\emph{Result 2:}} The frequentist expectations of $R_{post}(
\mbox{\boldmath $x$})$ and $1/R_{post}(\mbox{\boldmath $x$})$ over the
rejection region are
\[
E_{\mbox{\boldmath $x$}}[R_{post}(\mbox{\boldmath $x$})\mid H_{0},\mathcal{R}]=R_{pre}\quad
\mbox{and}\quad E_{\mbox{\boldmath $x$}}[1/R_{post}(\mbox{\boldmath $x$})\mid H_{1}^{\ast },
\mathcal{R}]=[R_{pre}]^{-1}\,,
\]
where $H_{1}^{\ast }$ refers to the marginal alternative model with density $
m(\mbox{\boldmath $x$})$ (defined in (\ref{eq.marginal})).

\noindent \emph{Proof.} See Appendix.

The first identity states that, under $H_{0}$, the average of the
post-experimental rejection ratios over the rejection region when rejecting\
(the `long-run average reported accuracy') equals the pre-experimental
rejection ratio (the `long-run average actual accuracy'). Hence, the frequentist
principle is satisfied: if a frequentist reports Bayes factors, then the long-run
average will be the pre-experimental rejection ratio.

The second identity is an analogous result that holds under $H_{1}$.
Whereas the pre- and post-experimental rejection ratios relate to the
relative likelihood of $H_{1}$ to $H_{0}$, the reciprocal of these
quantities relate to the relative likelihood of $H_{0}$ to $H_{1}$.
This identity states that the long-run average of the reciprocal of
the post-experimental rejection ratio will be the reciprocal of
the pre-experimental rejection ratio.

The first identity
is completely frequentist, as it involves only the density of the data under
the null hypothesis. The second identity, however,
is not strictly frequentist because it involves the marginal density
of the data (i.e., averaging over all possible non-null
values of $\theta $ in addition to averaging over the data); the long-run
average behavior of $R_{post}(
\mbox{\boldmath $x$})$ if $\theta $ is not null would not be its
behavior averaged across values of $\theta $, but rather its behavior
under the true value of $\theta $. For this reason, the discussion hereafter
focuses on the first identity.

\medskip \noindent \textbf{Example 4 (continued):} \emph{Effectiveness of
An AIDS Vaccine}: To illustrate the first identity in Result 2, Figure 4
presents $R_{post}(z)$ as a function of $z$ over the
rejection region for Example 4. The value of $R_{post}(z)$
itself ranges from $2:1$ (for data at the boundary of the rejection region)
to $\infty $. The weighted average of $R_{post}(z)$ as $z$ ranges from $1.645$
to $\infty $ (the rejection region)\ is $9:1$ (weighted with respect to the
density of $z$ under the null hypothesis). If one observed $z=1.645$ (a $p$-value of
0.05) or the actual $z=2.06$ (a $p$-value of 0.02), the pre-experimental
rejection ratio of $9:1$ would be an overstatement of the actual rejection
ratio; if, say, instead, $z=3$ had been observed, the post-experimental
rejection ratio would be $35:1$, much larger than the pre-experimental
rejection ratio.\bigskip

\begin{figure}[tbp]
\par
\begin{center}
\includegraphics[scale=0.55]{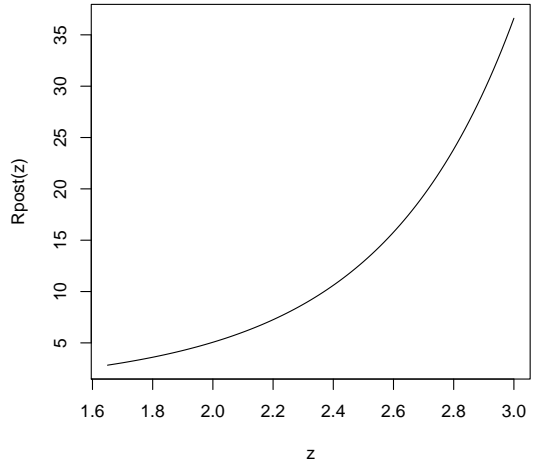} \label{fig:BF}
\end{center}
\par
\vspace{-2em}
\caption{In the vaccine example, $R_{post}(z)$ as a function of $z$ over the
rejection region.}
\end{figure}

The possibility that the data could generate post-experimental rejection
ratios larger or smaller than the pre-experimental rejection ratio is a
logical necessity, as the latter is an average of the former. Thus
logically, the post-experimental rejection ratios must be smaller than the
pre-experimental rejection ratio for data near the critical value of the
rejection region, with the reverse being true for data far from the critical
value.

This frequentist/Bayes equivalence also works for any composite null
hypothesis that has a suitable invariance structure,\footnote{Here, {\em invariance}
refers to a mathematical theory --- concerning transformations of data
and parameters in models --- that applies when the transformed model
has the same structure as the original; see Dass and Berger (2003) for
formal definition within the context of this discussion.} as long as the alternative
hypothesis also shares the invariance structure. As a simple example, it
would apply to the very common case of testing the null hypothesis that a normal mean is zero,
versus the alternative that it is not zero, when the normal model has an
unknown variance. Classical testing in this situation can be viewed as
reducing the data to consideration of the $t$-statistic and the noncentral $
t $-distribution and testing whether or not the mean of this distribution is
zero. As the null hypothesis here is now simple, the above result applies.
(More generally, the equivalence follows by reducing the data to what is
called the `maximal invariant statistic'\, for which the null hypothesis
becomes simple; see Berger, Boukai and Wang, 1997, for this reduction in the
above example, and Dass and Berger, 2003, for the general invariance theory.)

Result 2 raises a philosophical question: \emph{How can }$R_{post}(\mbox{\boldmath $x$})$
\emph{\ be a frequentist procedure if it depends on a prior distribution?}
The answer is that $R_{post}(\mbox{\boldmath $x$})$ defines a class of
optimal frequentist procedures, indexed by prior distributions. Each prior
distribution $\pi (\theta )$ results in a procedure whose post-experimental
rejection ratio equals the pre-experimental rejection ratio in expectation
(and, hence, is a valid frequentist procedure); different priors simply
induce different power characteristics.

Indeed, $R_{post}(\mbox{\boldmath $x$})$ will tend to be large if the
alternative is true and the true value of $\theta$ is where $\pi(\theta)$
predicts it to be. Thus, for a frequentist, $\pi(\theta)$ can simply be viewed as a device to
optimally power the procedure in desired locations. Note that these
locations need not be where $\theta$ is believed to be. For instance, a
common criterion in classical design is to select a value $\theta_1$ in the
alternative that is viewed as being `practically different from $\theta_0$'\
in magnitude, and then designing the experiment to have significant power at
$\theta_1$. For $R_{post}(\mbox{\boldmath $x$})$, one could similarly choose
the prior to be centered around $\theta_1$ (or, indeed, to be a point
mass at $\theta_1$).

The pre-experimental rejection ratio $R_{pre}$ is a frequentist
rejection-ratio procedure that does not depend on the data; it
effectively reports $R_{post}(\mbox{\boldmath $x$})$ to be a constant (e.g.,
imagine the constant line at 9 in Figure 4). This procedure, however, is not
obtainable from any prior distribution. The reason is that it is the \emph{uniformly worst}
procedure, when examined from a conditional frequentist perspective. The
intuition behind this claim can be seen from Figure 4. Any curve that has
the right frequentist expectation is a valid frequentist report. Curves that
are decreasing in $z$ would be nonsensical (reporting lower rejection ratios
the more extreme the data), so the candidate curves are the nondecreasing
curves. The constant curve (i.e., the pre-experimental rejection ratio) is
the worst of this class, as it makes no effort to distinguish between data
of different strengths of evidence.

While Result 2 shows that a frequentist is as entitled to report $R_{post}(\mbox{\boldmath $x$})$
as to report $R_{pre}$, the logic just outlined shows that $R_{post}(\mbox{\boldmath $x$})$ is
clearly a superior frequentist report to $R_{pre}$, as it is reflective of
the strength of evidence in the actual data, rather than an average of all
possible data in the rejection region.\bigskip

\noindent \textbf{Example 3 (continued):} \emph{Genome-Wide Association
Studies}: Recall the above example of the Wellcome Trust Case Control
Consortium, who explicitly calculated the pre-experimental rejection ratio
in order to justify their significance threshold. The article also reported
the Bayes factors, $R_{post}(\mbox{\boldmath $x$})$, for their 21
discoveries, and the post-experimental odds $O_{post}=O_{P}\times R_{post}(
\mbox{\boldmath
$x$})$. These ranged between $\frac{1}{10}$ and $10^{68}$ for the 21 claimed
associations. Thus the post-experimental odds ranged from 1 to 10 against $
H_{1}$ to overwhelming odds in favor of $H_{1}$; reporting these
data-dependent odds seems much preferable to always reporting 10 to 1,
especially because reporting $R_{post}(\mbox{\boldmath$x$})$ is every bit as
frequentist as reporting the pre-experimental rejection ratio. \bigskip

\section{Choosing the Priors for the Post-Experimental Odds}

A potential objection to using the post-experimental rejection
odds is that it requires choosing priors: the prior odds $\frac{\pi _{1}}{\pi _{0}}$
and the prior distribution $\pi (\theta )$ for $\theta$ under
the alternative hypothesis. In the previous section, we addressed
philosophical objections to the latter, and we showed that it has
fully frequentist justification. In this section, we address the
practical question: which priors should a researcher choose?

For the prior odds $\frac{\pi _{1}}{\pi _{0}}$,
one option is to report conclusions for a range of plausible prior odds.
Another option is to focus the analysis entirely on the Bayes factor, without taking
a stand on the prior odds. A Bayesian reader can easily apply his or her
own prior odds to draw conclusions.

For choosing the prior distribution $
\pi (\theta )$, here are some options:

\medskip \noindent \textbf{1. \ Subjective prior:} When a subjective prior
is available, such as the `study team prior'\ in the Example 4, using it is
optimal. Again, note that the resulting procedure is still a frequentist
procedure with the prior just being used to tell the procedure where high
power is desired (as discussed above).

\medskip \noindent \textbf{2. \ Power considerations:} If the experiment was
designed with power considerations in mind, use the implicit prior that was
utilized to determine power. This could be a weight function (the same
thing as a prior density, but a language preferred by frequentists) if used
to compute power, or a specified point (i.e., a prior giving probability one
to that point) if that is what was done.

\medskip \noindent \textbf{3. \ Objective Bayes conventional priors:}
Discussion of these can be found in Berger and Pericchi (2001). One popular
such prior, that applies to our testing problem, is the \emph{intrinsic prior}
defined as follows:

\begin{itemize}
\item Let $\pi ^{O}(\theta )$ be a good estimation objective prior (often a
constant), with resulting posterior distribution and marginal distribution
for data $\mbox{\boldmath $x$}$ given, respectively, by
\[
\pi ^{O}(\theta \mid \mbox{\boldmath $x$})=f(\mbox{\boldmath $x$}\mid \theta
)\pi ^{O}(\theta )/m^{O}(\mbox{\boldmath $x$}),\quad m^{O}(
\mbox{\boldmath
$x$})=\int f(\mbox{\boldmath $x$}\mid \theta )\pi ^{O}(\theta )\,d\theta \,.
\]

\item Then the intrinsic prior (which will be proper) is
\[
\pi ^{I}(\theta )=\int \pi ^{O}(\theta \mid \mbox{\boldmath $x$}^{\ast })f(
\mbox{\boldmath $x$}^{\ast }\mid \theta _{0})\,d\mbox{\boldmath $x$}^{\ast
}\,,
\]
with $\mbox{\boldmath $x$}^{\ast }=(x_{1}^{\ast },\ldots ,x_{q}^{\ast })$
being imaginary data of the smallest sample size $q$ such that $m^{O}(
\mbox{\boldmath $x$}^{\ast })<\infty $.
\end{itemize}

$\pi ^{I}(\theta )$ is often available in closed form, but even if not,
computation of the resulting Bayes factor is often a straightforward
numerical exercise.

\medskip \noindent \textbf{4. \ Empirical Bayes prior:} This is found by
maximizing the numerator of $R_{post}(\mbox{\boldmath $x$})$ over some class
of possible priors. Common are the class of nonincreasing priors away from $
\theta _{0}$ or even the class of all priors; both were considered in
Example 4.

\medskip \noindent \textbf{5. }\ $p$\textbf{-value bound:} Instead of
picking a prior distribution to calculate $R_{post}(\mbox{\boldmath $x$})$,
use the generic upper bound on $R_{post}(\mbox{\boldmath $x$})$ that was
discussed in Section 3.2.

\medskip \noindent \textbf{Evaluation of these methods:} Any of the first
three approaches are preferable because they are logically coherent, from
both a frequentist and Bayesian perspective. Option 1 is clearly best if
either beliefs or power considerations allow for the construction of the
prior distribution or power `weight function'. Note that there is no issue
of `subjectivity' versus `objectivity' here, as this is still a fully
frequentist procedure; the prior/power-weight-function is simply being used
to `place your frequentist bets' as to where the effect will be. (We
apologize to Bayesians who will be offended that we are not separately
dealing with prior beliefs and `effect sizes' that should enter through a
utility function; we are limited by the scope of our paper.)

Option 2 (the power approach)\ is the same as Option 1 if a `weight
function' approach to power was used. If `power at a point' was done in
choosing the design, one is facing boom or bust. If the actual effect size
is near the point chosen, the researcher will have maximized
post-experimental power; otherwise, one may be very underpowered to detect a
clear effect.

Option 3 (the intrinsic prior approach) is highly attractive if either of
the first two approaches cannot be implemented. There is an extensive
literature discussing the virtues of this approach (see Berger and Pericchi,
2001 and 2015, for discussion and other references).

The last two approaches above suffer from two problems. First, they are
significantly biased (in the wrong way) from both Bayesian and frequentist
perspectives. Indeed, if ${\bar{R}_{post}}(\mbox{\boldmath $x$})$ is the
answer obtained from either approach, then
\[
R_{post}(\mbox{\boldmath $x$})<{\bar{R}_{post}}(\mbox{\boldmath $x$}
)\,,\quad \quad R_{pre}<E[{\bar{R}_{post}}(\mbox{\boldmath $x$})\mid H_{0},
\mathcal{R}]\,.
\]
Thus, in either case, one is reporting larger rejection ratios in favor of $
H_{1}$ than is supported by the data.

The (hopefully transient) appeal of using the last two approaches is that
they are easy to implement --- especially the last. And the answers, even if
biased in favor of the alternative hypothesis, are so much better than
$p$-values that their use would significantly improve science.

In short, the practical problem of choosing the prior distribution $
\pi (\theta )$ is \emph{not} a compelling argument against the
post-experimental odds approach. There are a range of options available,
depending on the context and the researchers' comfort level with statistical
modeling. For example, Option 5 is simple and doable in essentially every
context, as it avoids the need to specify $
\pi (\theta )$ altogether.

\section{Post-Experimental Rejection Ratios Are\ Immune to Optional Stopping}

A common practice in psychology is to ignore optional stopping (John,
Loewenstein, and Prelec, 2012): if one is close to $p=0.05$, go get more
data to try get below 0.05 (with no adjustment).\bigskip

\noindent \textbf{Example 6:} \emph{Optional Stopping}: Suppose one has $
p=0.08$ in a sample of size $n$ in testing whether or not the mean is
zero. And suppose it is known that the data are drawn from a normal distribution
with known variance. If one sequentially takes up to four additional
samples of size $\frac{n}{4}$, computing the $p$-value (without adjustment)
for the accumulated data at each stage, an easy computation shows that the
probability of reaching $p=0.05$ at one of the four stages (at which point
one would stop) is $\frac{2}{3}$. Thus, when using $p$-values to assess
significance, optional stopping is cheating. By ignoring optional stopping,
one has a large probability of getting to `significance.'\ (Indeed, if one
kept on taking additional samples and computing the $p$-value with no
adjustment, one would be guaranteed of reaching $p=0.05$ eventually, even
when $H_{0}$ is true.)\bigskip

If one sequentially observes data $\mbox{\boldmath $x$}_{1},
\mbox{\boldmath
$x$}_{2},\ldots $ (where each $\mbox{\boldmath $x$}_{i}$ could be a single
observation or a batch of data), a \emph{stopping rule} $
\mbox{\boldmath
$\tau$}$ is a sequence of indicator functions $\mbox{\boldmath $\tau $}
=(\tau _{1}(\mbox{\boldmath $x$}_{1}),\tau _{2}(\mbox{\boldmath $x$}_{1},
\mbox{\boldmath $x$}_{2}),\ldots )$ which indicate whether or not
experimentation is to be stopped depending on the data observed so far. The
only technical condition we impose on the stopping rule is that it be
\emph{proper}: the probability of stopping eventually must be one.

\medskip
\noindent \textbf{Example 6 (continued):} \emph{Optional Stopping}: In the
example, $\mbox{\boldmath $x$}_{1}$ would be the original data of size $n$
and $\mbox{\boldmath $x$}_{2},\ldots ,\mbox{\boldmath $x$}_{5}$ would be the
possible additional samples of size $n/4$ each. The stopping rule would be
\begin{eqnarray*}
\tau _{1}(\mbox{\boldmath $x$}_{1}) &=&\left \{
\begin{array}{c}
1\quad \mbox{if $p(\mbox{\boldmath $x$}_1) < 0.05$} \\
0\quad \quad \quad \mbox{otherwise}
\end{array}
\,,\right. \\
\tau _{2}(\mbox{\boldmath $x$}_{1},\mbox{\boldmath $x$}_{2}) &=&\left \{
\begin{array}{c}
1\quad \mbox{if $p(\mbox{\boldmath $x$}_1, \mbox{\boldmath $x$}_2) < 0.05$}
\\
0\quad \quad \quad \quad \quad \mbox{otherwise}
\end{array}
\,,\right. \\
&\vdots & \\
\tau _{5}(\mbox{\boldmath $x$}_{1},\ldots ,\mbox{\boldmath $x$}_{5}) &=& 1
\,.
\end{eqnarray*}
An unconditional frequentist \emph{must} incorporate the stopping rule into
the analysis for correct evaluation of a procedure --- not doing so is really
no better than making up data. Thus, for a rejection region $\mathcal{R}$,
the frequentist type I error would be $\alpha =\Pr(\mathcal{R}\mid \theta _{0},
\mbox{\boldmath $\tau$})$, the probability being taken with respect to the
stopped data density
\[
\tau _{N}(\mbox{\boldmath $x$}_{1},\mbox{\boldmath $x$}_{2},\ldots ,
\mbox{\boldmath $x$}_{N})f(\mbox{\boldmath $x$}_{1},\mbox{\boldmath $x$}
_{2},\ldots ,\mbox{\boldmath $x$}_{N}\mid \theta_0 )\,,
\]
where $N$ denotes the (random) stage at which one stops. Power would be
similarly defined, leading to the rejection ratio $R_{pre}$, which will
depend on the stopping rule.

In contrast, it is well known (cf. Berger, 1985, and Berger and Berry, 1988)
that the Bayes factor \emph{does not} depend on the stopping rule. That is,
if the stopping rule specifies stopping after observing $(
\mbox{\boldmath
$x$}_{1},\ldots ,\mbox{\boldmath $x$}_{k})$, the Bayes factor computed using
the stopped data density will be identical to that assuming one had a
predetermined fixed sample $(\mbox{\boldmath $x$}_{1},\ldots ,
\mbox{\boldmath $x$}_{k})$. Intuitively, even though the stopping rule
will cause some data to be
especially likely to be observed --- in particular, data that causes the $p$-value
to just cross the significance threshold --- the likelihood of observing that
data is increased under both the null hypothesis and the alternative
hypothesis, leaving the likelihood ratio unaffected. In the formal
derivation of the result, the factor $\tau _{N}(\mbox{\boldmath $x$}_{1},\mbox{\boldmath $x$}_{2},\ldots ,
\mbox{\boldmath $x$}_{N})$ appears in both the numerator and
denominator of the Bayes factor and therefore cancels out.

There are two consequences of this result:

\begin{description}
\item[1.] Use of the Bayes factor gives experimenters the freedom to employ
optional stopping without penalty. (In fact, Bayes factors can be used in
the complete absence of a sampling plan, or in situations where the analyst
does not know the sampling plan that was used.)

\item[2.] There is no harm if `undisclosed optional stopping' is used, as
long as the Bayes factor is used to assess significance. In particular, it
is a consequence that an experimenter cannot fool someone through use of
undisclosed optional stopping.\bigskip
\end{description}

\textbf{Example 6 (continued):} \emph{Optional Stopping}: Suppose the study
reports a $p$-value of 0.05 and no mention is made of the stopping rule. A
conventional objective Bayesian analysis will result in a Bayes factor such
as $R_{post}=2$. This will certainly not mislead people into thinking the
evidence for rejection is strong.\bigskip

The frequentist/Bayesian duality argument from the previous section still
also holds, so that a frequentist can also report the Bayes
factor --- ignoring the stopping rule --- and it is a valid frequentist report.
That is, conditional on stopping within the rejection region, the reported
ratio of correct to incorrect rejection does not depend on the stopping
rule\footnote{
This result does \emph{not} apply to the Bayes factor bound in (\ref
{eq.bound}), however. That bound assumed that the $p$-value is proper, which
does not hold if optional stopping is ignored in its computation.}. This is
remarkable and seems like cheating, but it is not. (See Berger, 1985, and
Berger and Berry, 1988, for much more extensive discussion concerning this
issue.)

To be sure, a frequentist would still need to determine the rejection region
$\mathcal{R}$ so as to achieve desired Type I and Type II errors and the
implied (pre-experimental) rejection ratio. And if one reads an article in
which optional stopping was utilized and not reported, one cannot be sure
what rejection region was actually used, and so one cannot calculate the
pre-experimental rejection ratio. But these are minor points as long as post-experimental
rejection ratios are reported; as they do not depend on the
stopping rule, the potential to mislead is dramatically reduced.

\section{Summary: Our Proposal for Statistical Hypothesis Testing of Precise
Hypotheses}

Our proposal can be boiled down to two recommendations: report the
pre-experimental rejection ratio when presenting the experimental design,
and report the post-experimental rejection ratio when presenting the experimental results.
These recommendations can be implemented in a range of ways, from
full-fledged Bayesian inference to very minor modifications
of current practices. In this section, we flesh out the range of
possibilities for each of these recommendations,
drawing on the points discussed throughout this paper,
in order from smallest to largest deviations from current practice.

\medskip \noindent \emph{1. Report the
pre-experimental rejection ratio when presenting the experimental design.}
This recommendation can be carried out in any research community that is
comfortable with power calculations. Reporting the pre-experimental rejection ratio
 --- the ratio of power to Type I error --- is a wonderful way to summarize the expected persuasiveness
of any significant results that may come out from the experiment.

Of course, calculating power prior to running an experiment has long been
part of recommended practice. Our emphasis is on the usefulness
of such calculations in ensuring that statistically significant
results will constitute convincing evidence. Moreover, beyond conducting power
calculations, \emph{reporting} them and the anticipated effect sizes can
help `keep us honest'\ as researchers:\ knowing that we are accountable to
skeptics and critically-minded colleagues encourages us to keep our
anticipated effect sizes realistic rather than optimistic.\footnote{
Such reporting would have the additional advantage of facilitating habitual
discussion of how the observed effect sizes compared to those that were
anticipated and those obtained in related work. Doing so provides information
about the plausibility of the observed effect. For example, if the observed
effect size is much larger than anticipated, the researcher might be
prompted to search for a potential confound that could have generated
the large effect.}

Moving further away from current practice in many disciplines, we
recommend that researchers report their prior odds (for the alternative hypothesis relative to
the null hypothesis), or a range of reasonable prior odds.\footnote{
In addition to reporting their own priors, researchers could report
the priors of other researchers. For example, it might be
useful to report the results of surveying colleagues about
what results they expect from the experiment. More ambitiously,
prediction markets could be used to aggregate the beliefs of many
researchers (Dreber et al., 2015).} Research that
convincingly verifies surprising predictions of a theory is a major advance
and deserves to be more famous and better published. But when the
predictions are surprising --- that is, when the prior odds in favor of the
alternative hypothesis are low --- the evidence should have to be more
convincing before it is sufficient to overturn our skepticism. That is, when the prior
odds are lower, the pre-experimental rejection ratio should be required
to be higher in order for the experimental design to be deemed appropriate.
At a fixed significance threshold of 0.05, achieving a higher pre-experimental
rejection ratio requires running a higher-powered experiment.

To further improve current practice, Type I error of 0.05
should not be a one-size-fits-all significance threshold. Null
hypotheses that have higher prior odds should be required to reach
a more stringent significance threshold before they are rejected.
Given the researchers' prior odds, the
appropriate significance threshold can be calculated easily as described in
Section 2.2.

\medskip \noindent \emph{2. Report the
post-experimental rejection ratio (Bayes factor) when presenting the experimental results.}
After seeing the data,
the pre-experimental rejection ratio should be replaced by its post-experimental counterpart,
the Bayes factor: the likelihood of the observed data under the alternative hypothesis
relative to the likelihood of the observed data under the null hypothesis.
This measure of the strength of the evidence has full frequentist
justification and is much more accurate than the pre-experimental measure.

The simplest version of this recommendation is to report the Bayes factor
bound: $1/[-ep\log (p)]$. Calculating this bound is simple because the only
input is the $p$-value obtained from any standard statistical test.
Although it only gives an upper bound on what a
$p$-value means in terms of the post-experimental ratio of correct to incorrect
rejection of the null hypothesis, it is reasonably accurate for
well powered experiments. By alerting researchers when seemingly strong evidence
is actually not very compelling, reporting of the Bayes factor bound
would go far by itself in improving interpretation of experimental results.

Even better is to calculate the actual Bayes factor, although doing so requires
some statistical modeling (as illustrated in Example 4) and specification of a `prior distribution' of the
effect size under the alternative hypothesis, $\pi (\theta )$. The
frequentist interpretation of $\pi (\theta )$ is as a `weight function' that
specifies where it is desired to have high power for detection of an effect.
Hence, if power calculations were used in the experimental design, then the
effect size (or distribution of effect sizes) used for the power calculation
can be used directly as $\pi (\theta )$. Other possible `objective' choices
for $\pi (\theta )$ that are well developed in the statistics literature
include the intrinsic prior or an empirical Bayes prior.

In research communities in which subjective priors are acceptable, then (as
we also recommended in the context of the pre-experimental rejection ratio)
researchers should report their prior odds, or a reasonable range,
and draw conclusions in light of both the evidence and the prior
odds. Indeed, among all of our recommendations, our `top pick'\
would be to report results in
terms of the post-experimental odds of the hypotheses: the product of the
prior odds (which may be highly subjective) and the Bayes factor
(which is much less subjective). Researchers comfortable with subjective
priors could also choose a subjective prior for $\pi (\theta )$. Of course,
to the extent possible, subjective priors --- like anticipated effect sizes
more generally --- should be justified with reference
to what is known about the phenomenon under study and about related phenomena,
taking into account publication and other biases.

\bigskip

Many of the key parameters relevant for interpreting the evidence --- such as
anticipated effect sizes, the significance threshold, the pre-experimental
rejection ratio, and the prior odds --- should be possible to set prior to running the
experiment. We therefore further recommend preregistering these parameters. As many
have argued, preregistration would help researchers to avoid the hindsight
bias (Fischhoff, 1975), not to mention any temptation to tweak the
parameters ex post, and hence would make the data analysis more credible.\footnote{
Many versions of preregistration have been proposed, ranging from researcher-initiated
pre-analysis plans that may be deviated from (e.g., Olken, 2015), to journal-enforced
preregistration of experimental designs that are peer-reviewed prior to running
the experiment (e.g., Chambers et al., 2014). There have also been
many thoughtful criticisms of preregistration requirements that may be too rigid or too
onerous (e.g., Coffman and Niederle, 2015). Addressing what form
preregistration should take and how strictly it should be required or enforced
is beyond the scope of this paper.}


\subsection*{Appendix 1: Proof of Result 2}

We first prove the result for the expected value under the null.
Since
\[
f(\mbox{\boldmath $x $}\mid H_{0},\mathcal{R})=\, \frac{f(
\mbox{\boldmath
$x $}\mid \theta _{0})}{\mbox{Pr}(\mathcal{R}\mid H_{0})}\, \qquad
\mbox{if \
}\mbox{\boldmath $x $}\in \mathcal{R}\ ,
\]
and $0$ otherwise, it follows that
\begin{eqnarray*}
E[R_{post}(\mbox{\boldmath $X$}) &\mid &H_{0},\mathcal{R}]=\int \frac{m(
\mbox{\boldmath $x$})}{f(\mbox{\boldmath
$x $}\mid \theta _{0})}\,f(\mbox{\boldmath $x $}\mid H_{0},\mathcal{R})d
\mbox{\boldmath $x $} \\
&=&\, \frac{1}{\mbox{Pr}(\mathcal{R}\mid H_{0})}\, \int_{\mathcal{R}}m(
\mbox{\boldmath $x$})\,d\mbox{\boldmath $x $} \\
&=&\, \frac{1}{\alpha }\, \int_{\Theta }\left[ \int_{\mathcal{R}}f(
\mbox{\boldmath $x $}\mid \theta )\,d\mbox{\boldmath $x  $}\right] \pi
(\theta )\,d\theta
\end{eqnarray*}
and the result follows trivially by noting that $\int_{\mathcal{R}}f(
\mbox{\boldmath $x $}\mid \theta )\,d\mbox{\boldmath $x $}=1-\beta (\theta )$
.

Under the alternative, we consider the testing of $H_{0}:
\mbox{\boldmath $X
$}\sim f(\mbox{\boldmath $x $}\mid \theta _{0})\quad \mbox{vs.}\quad
H_{1}^{\ast }:\mbox{\boldmath $X $}\sim m(\mbox{\boldmath $x$})$, so now
\[
f(\mbox{\boldmath $x $}\mid H_{1}^{\ast },\mathcal{R})=\, \frac{f(
\mbox{\boldmath $x $}\mid H_{1}^{\ast })}{\mbox{Pr}(\mathcal{R}\mid
H_{1}^{\ast })}\,=\, \frac{m(\mbox{\boldmath $x$})}{\int_{\mathcal{R}}m(
\mbox{\boldmath $x$})\,d\mbox{\boldmath $x$}}\qquad \mbox{if \ }
\mbox{\boldmath $x $}\in \mathcal{R}\ ,
\]
and $0$ otherwise; now, as shown above ${\int_{\mathcal{R}}m(
\mbox{\boldmath
$x$})\,d\mbox{\boldmath $x$}}=1-\bar{\beta}$, so that
\[
E[1/R_{post}(\mbox{\boldmath $x$})\mid H_{1}^{\ast },\mathcal{R}]=\, \frac{1
}{1-\overline{\beta }}\, \int_{\mathcal{R}}\frac{1}{R_{post}(
\mbox{\boldmath
$x$})}\,m(\mbox{\boldmath $x$})\,d\mbox{\boldmath $x  $}=\, \frac{1}{1-
\overline{\beta }}\, \int_{\mathcal{R}}f(\mbox{\boldmath $x $}\mid \theta
_{0})\,d\mbox{\boldmath $x $}
\]
which gives the desired result.


\begin{thebibliography}{Wellcome Trust Case Control Consortium, 2007}
\bibitem[Anscombe, 1954]{Ansc:54} Anscombe, F.J. (1954). Fixed-Sample-Size
Analysis of Sequential Observations. \textit{Biometrics }\textbf{10}(1),
89-100.

\bibitem[Bayarri et al., 2012]{BayaBergEtAll:12} Bayarri, M.J., Berger,
J.O., Forte, A., Garc\`ia-Donato, G. (2012). Criteria for Bayesian model
choice with application to variable selection. \textit{Annals of Statistics}
\textbf{40}, 1550-1577.

\bibitem[Bayarri et al., 2013]{BayaBerg:13} Bayarri, M.J. and Berger, J.
(2013). Hypothesis testing and model uncertainty. In \textit{Bayesian Theory
and Applications}, Paul Damien, Petros Dellaportas, Nicholas Polson and
David Stephens (Eds.), Oxford University Press, 361--400.

\bibitem[Bem, 2011]{Bem:11} Bem, D.J. (2011). Feeling the Future:
Experimental Evidence for Anomalous Retroactive Influences on Cognition and
Affect. \textit{Journal of Personality and Social Psychology} \textbf{100}%
(3), 407-25.

\bibitem[Benjamin at al., 2010]{BenjChoiStri:10} Benjamin, D.J., Choi, J.J.,
Strickland, A.J. (2010). Social Identity and Preferences. \textit{American
Economic Review} \textbf{100}, 1913--1928.

\bibitem[Berger, 1985]{Berg:85} Berger, J. (1985). \textit{Statistical Decision Theory and
Bayesian Analysis}, Springer--Verlag, New York, 1985.

\bibitem[Berger, 2003]{Berg:03} Berger, J. (2003). Could Fisher, Jeffreys
and Neyman have agreed on testing (with Discussion)? \textit{Statistical
Science}, \textbf{18}, 1--32.

\bibitem[Berger and Berry, 1988]{BergBerr:88} Berger, J. and Berry, D.
(1988). The relevance of stopping rules in statistical inference (with
Discussion). In \textit{Statistical Decision Theory and Related Topics IV.\/}
Springer--Verlag, New York.

\bibitem[Berger et al., 1997]{BergBoukWang:97} Berger, J., Boukai, B. and
Wang, Y. (1997). Unified frequentist and Bayesian testing of a precise
hypothesis (with discussion). \textit{Statistical Science}, \textbf{12(3)},
133--160.

\bibitem[Berger et al., 1999]{BergBoukWang:99} Berger, J., Boukai, B. and
Wang, Y. (1999). Simultaneous Bayesian-frequentist sequential testing of
nested hypotheses. \textit{Biometrika}, \textbf{86}, 79--92.

\bibitem[Berger et al., 1994]{BergBrowWolp:94} Berger, J., Brown, L. and
Wolpert, R. (1994). A unified conditional frequentist and Bayesian test for
fixed and sequential hypothesis testing. \textit{Ann. Statist.} \textbf{22},
1787-1807.

\bibitem[Berger et al., 1999]{BergMort:99} Berger, J. and Mortera, J.
(1999). Default Bayes factors for non-nested hypothesis testing. \textit{J.
Amer.\ Statist.\ Assoc.\/}, \textbf{94}, 542--554.

\bibitem[Berger et al., 2001]{BergPerr:01} Objective Bayesian methods for
model selection: introduction and comparison (with Discussion). In \textit{%
Model Selection}, P. Lahiri, ed., Institute of Mathematical Statistics
Lecture Notes -- Monograph Series, volume 38, Beachwood Ohio, 135--207.

\bibitem[Berger and Pericchi, 2015]{BergPerr:15} Berger, J. and Pericchi, L.
(2015). Bayes Factors. \textit{Wiley StatsRef: Statistics Reference Online}
1--14.

\bibitem[Brown, 1978]{Brow:78} Brown, L. D. (1978). A contribution to
Kiefer's theory of conditional confidence procedures, \textit{Annals of
Statistics}, \textbf{6}, 59-71.

\bibitem[Button et al., 2013]{ButtIoanMokrNoseFlinRobiMuna:13} Button, K.S.,
Ioannidis, J.P.A., Mokrysz, C., Nosek, B.A., Flint, J., Robinson, E.S.J.,
Munaf\`{o}, M.R. (2013). Power failure: why small sample size undermines the
reliability of neuroscience. \textit{Nature Reviews Neuroscience} \textbf{14},
365-376.

\bibitem[Chambers et al., 2014]{Chambers:14} Chambers, C.,
Feredoes, E., Muthukumaraswamy, S.D., Etchells, P. (2014). Instead of `Playing
the Game’ it is Time to Change the Rules: Registered Reports at
AIMS Neuroscience and Beyond. \textit{AIMS Neuroscience}, \textbf{1}, 4–-17.

\bibitem[Coffman, 2015]{CoffNied:15} Coffman, L.C., Niederle, M. (2015). Pre-Analysis 
Plans Have Limited Upside Especially Where Replications Are Feasible. 
\textit{Journal of Economic Perspectives}, \textbf{29(3)}, 81–-98.

\bibitem[Cohen, 1988]{Cohen:88} Cohen, J. (1988). \textit{Statistical power
analysis for the behavioral sciences}, Erlbaum, Hillsdale, NJ, 1988.


\bibitem[Dass, 2003]{DassBerg:03} Dass, S. and Berger, J. (2003). Unified
Bayesian and conditional frequentist testing of composite hypotheses.
\textit{Scandinavian Journal of Statistics}, \textbf{30}, 193--210.

\bibitem[Dreber et al., 2015]{Dreber:15} Dreber, A.,
Pfeiffer, T., Almenberg, J., Isaksson, S., Wilson, B., Chen, Y., Nosek, B.,
Johannesson, M. (2015). Using Prediction Markets to Estimate the
Reproducibility of Scientific Research. \textit{Proceedings of the National
Academy of Sciences of the United States of America}, \textbf{112(50)}, 15343–-15347.


\bibitem[Edwards, Lindman, and Savage, 1963]{Edwards:63} Edwards, W., Lindman, H.,
and Savage, L. (1963).
Bayesian Statistical Inference for Psychological Research. \textit{%
Psychological Review,} \textbf{70(3)}, 193–-242.

\bibitem[Fischhoff, 1975]{Fisc:75} Fischhoff, B. (1975). Hindsight $\neq $
foresight. The effect of outcome knowledge on judgment under uncertainty.
\textit{Journal of Experimental Psychology: Human Perception and Performance}
\textbf{1}(3), 288-299.

\bibitem[Fischhoff, 1983]{Fisc:83} Fischhoff, B., Beyth-Marom, R. (1983).
Hypothesis Evaluation From a Bayesian Perspective. \textit{Psychological
Review} \textbf{90}(3), 239-260.

\bibitem[Garner, 2007]{Garn:07} Garner, C. (2007). Upward bias in odds ratio
estimates from genome-wide association studies. \textit{Genetic Epidemiology}
\textbf{31}, 288--295.

\bibitem[Gelman, 2014]{Gelm:14} Gelman, A., Carlin, J. (2014). Beyond Power
Calculations: Assessing Type S (Sign) and Type M (Magnitude) Errors. \textit{%
Perspectives on Psychological Science} \textbf{9}(6), 641--651.

\bibitem[Gilbert et al., 2011]{Gilb:11} Gilbert, P., Berger, J., Stablein,
D., Becker, S., Essex, M., Hammer, S., Kim, J., and DeGruttola, V. (2011).
Statistical interpretation of the RV144 HIV vaccine efficacy trial in
Thailand: A case study for statistical issues in efficacy trials. \textit{%
The Journal of Infectious Diseases,} \textbf{203(7)}, 969--975.

\bibitem[Ioannidis, 2005]{Ioan:05} Ioannidis, J.P.A. (2005). Why Most
Published Research Findings Are False. \textit{PLoS Medicine} \textbf{2}(8),
124.

\bibitem[Ioannidis, 2008]{Ioan:08} Ioannidis, J.P.A. (2008). Effect of
formal statistical significance on the credibility of observational
associations. \textit{American Journal of Epidemiology} \textbf{168}(4),
374--383.

\bibitem[John et al., 2012]{JohnLoewPrel:12} John, L.K., Loewenstein, G.,
Prelec, D. (2012). Measuring the Prevalence of Questionable Research
Practices with Incentives for Truth-telling. \textit{Psychological Science}
\textbf{23}(5), 524--532.

\bibitem[Johnson, 2013]{John:13} Johnson, V. (2013). Revised standards for
statistical evidence. \textit{Proceedings of the National Academy of Sciences%
}, \textbf{110.48}, 19,313--19,317.

\bibitem[Kiefer, 1977]{Kief:77} Kiefer, J. (1977). Conditional confidence
statements and confidence estimators (with discussion). \textit{Journal of
the American Statistical Association}, \textbf{72}, 789-827.

\bibitem[Kruschke, 2011]{Krus:11} Kruschke, J. K. (2011). Bayesian
assessment of null values via parameter estimation and model comparison.\
\textit{Perspectives on Psychological Science} \textbf{6}(3), 299-312.

\bibitem[Locke et al., 2015]{LockEtAl:15} Locke, A.E., et al. (2015).
Genetic studies of body mass index yield new insights for obesity biology.
\textit{Nature} \textbf{518}, 197--206.

\bibitem[Lucas, 2000]{Luca:00} Lucas, M. (2000). Semantic priming without
association: A meta-analytic review. \textit{Psychonomic Bulletin and Review}
\textbf{6}, 618--630.

\bibitem[Lucke, 2009]{Luck:09} Lucke, J.F. (2009). A Critique of the
False-Positive Report Probability. \textit{Genetic Epidemiology} \textbf{33}%
, 145--150.

\bibitem[Masson, 2011]{Mass:11} Masson, M.E.J. (2011). A tutorial on a
practical Bayesian alternative to null-hypothesis significance testing.
\textit{Behavioral Research} \textbf{43}, 679--690.

\bibitem[Neyman, 1977]{Neym:77} Neyman, J. (1977). Frequentist probability
and frequentist statistics. \textit{Synthese} \textbf{36}, 97--131.

\bibitem[Olken, 2015]{Olken:15} Olken, B. (2015). Promises and Perils of
Pre-analysis Plans. \textit{Journal of Economic Perspectives} \textbf{29(3)}, 61–-80.


\bibitem[Richard et al., 2003]{RichBondStok:03} Richard, F.D., Bond, Jr.,
C.F., Stokes-Zoota, J.J. (2003). One Hundred Years of Social Psychology
Quantitatively Described. \textit{Review of General Psychology} \textbf{7}%
(4), 331--363.

\bibitem[Rietveld et al., 2013]{RietEtAl:13} Rietveld, C.A., et al. (2013).
GWAS of 126,559 individuals identifies genetic variants associated with
educational attainment.\  \textit{Science} \textbf{340}(6139), 1467--71.

\bibitem[Rietveld et al., 2014]{RietEtAl:14} Rietveld, C.A., et al. (2014).
Replicability and Robustness of Genome-Wide-Association Studies for
Behavioral Traits. \textit{Psychological Science} \textbf{25}(11),
1975--1986.

\bibitem[Ripke et al., 2014]{RipkEtAl:14} Ripke, S., et al. (2014).
Biological insights from 108 schizophrenia-associated genetic loci. \textit{%
Nature} \textbf{511}, 421--427.

\bibitem[Rosenthal, 1979]{Rose:79} Rosenthal, R. (1979). The `file drawer
problem’ and tolerance for null results. \textit{Psychological Bulletin}
\textbf{86}(3), 838-641.

\bibitem[Schkade, 1998]{Schk:98} Schkade, D.A., Kahneman, D. (1998). Does
Living in California Make People Happy? A Focusing Illusion in Judgments of
Life Satisfaction. \textit{Psychological Science}, \textbf{9}(5), 340-346.

\bibitem[Sellke, 1977]{Sell:12} Sellke, T.M. (2012). On the interpretation
of $p$-values, \textit{Tech. Rep. Department of Statistics, Purdue University%
}.

\bibitem[Sellke et al., 2001]{SellBayaBerg:12} Sellke, T., Bayarri, M.J.,
and Berger, J.O. (2001). Calibration of p Values for Testing Precise Null
Hypotheses, \textit{The American Statistician} \textbf{55}, 62-71.

\bibitem[Vankov et al., 2014]{VankBoweMuna:14} Vankov, I., Bowers, J., Munaf%
\`{o}, M.R. (2014). On the persistence of low power in psychological
science. \textit{Quarterly Journal Of Experimental Psychology} \textbf{67}%
(5), 1037-1040.

\bibitem[Visscher et al., 2012]{VissBrowMccaYang:12} Visscher, P.M., Brown,
M.A., McCarthy, M.I., Yang, J. (2012). Five years of GWAS discovery. \textit{%
American Journal of Human Genetics} \textbf{90}(1), 7-24.

\bibitem[Vovk, 1993]{Vovk:93} Vovk, V.G. (1993). A Logic of Probability,
with Application to the Foundations of Statistics. \emph{Journal of the
Royal Statistical Society. Series B}, \textbf{55}, 317--351

\bibitem[Wacholder et al., 2004]{WachEtAl:04} Wacholder, S., Chanock, S.,
Garcia-Closas, M., El ghormli, L., Rothman, N. (2004). Assessing the
Probability That a Positive Report is False: An Approach for Molecular
Epidemiology Studies. \textit{Journal of the National Cancer Institute}
\textbf{96}(6), 434-442.

\bibitem[Wagenmakers et al., in press]{WageEtAl:15} Wagenmakers, E.-J.,
Verhagen, A. J., Ly, A., Matzke, D., Steingroever, H., Rouder, J. N., Morey,
R. D. (in press). The need for Bayesian hypothesis testing in psychological
science. In Lilienfeld, S. O., \& Waldman, I. (Eds.), \textit{Psychological
Science Under Scrutiny: Recent Challenges and Proposed Solutions}. John
Wiley and Sons.

\bibitem[Wagenmakers et al., 2011]{WageEtAl:11} Wagenmakers, E.-J., Wetzels,
R., Borsboom, D., van der Maas, H. L. J. (2011). Why psychologists must
change the way they analyze their data: The case of psi: Comment on Bem
(2011). \textit{Journal of Personality and Social Psychology} \textbf{100},
426-432.

\bibitem[Wellcome Trust Case Control Consortium, 2007]{WellTrusEtAll:07} %
Wellcome Trust Case Control Consortium (2007). Genome-wide association study
of 14,000 cases of seven common diseases and 3,000 shared controls, \textit{%
Nature} \textbf{447}(7145), 661-678.

\bibitem[Wood, 2014]{WoodEtAl:14} Wood, A.R., et al. (2014). Defining the
role of common variation in the genomic and biological architecture of adult
human height. \textit{Nature Genetics} \textbf{46}, 1173--1186.

\bibitem[Zellner, 1986]{Zell:86} Zellner, A. (1986). On assessing prior
distributions and Bayesian regression analysis with g-prior distributions.
In {it Bayesian Inference and Decision Techniques: Essays in Honor of Bruno
de Finetti}, (eds. P. K. Goel and A. Zellner), 233-243.
North-Holland/Elsevier.

\bibitem[Zhang, 2013]{ZhanOrtm:13} Zhang, L., Ortmann, A. (2013). Exploring
the Meaning of Significance in Experimental Economics. Australian School of
Business Research Paper No. 2013-32.
http://papers.ssrn.com/sol3/papers.cfm?abstract\_id=2356018
\end{thebibliography}
\end{document}